\documentclass[aps,prd,twocolumn,nofootinbib,superscriptaddress]{revtex4-1}

\usepackage[utf8]{inputenc}

\usepackage{color}
\usepackage{graphicx}
\usepackage{amsmath}
\usepackage{amssymb}
\usepackage{bm}
\usepackage{acronym}
\usepackage{ifthen}
\usepackage{blindtext}
\usepackage[normalem]{ulem}
\usepackage{etoolbox}
\usepackage{xspace}
\usepackage{subfigure}
\usepackage{textcomp}
\usepackage{multirow}
\usepackage{lineno}
\usepackage{tabularx}
\usepackage{orcidlink}
\usepackage{siunitx}
\usepackage{hyperref}
\usepackage[capitalise]{cleveref}

\def\gwh{GW\xspace}
\def\gws{GWs\xspace}

\newacro{S/N}{signal-to-noise ratio}

\newacro{PN}{post-Newtonian}

\newacro{O3}{the third observing run}

\newacro{O2}{the second observing run}

\newacro{CW}{\emph{Continuous Wave}}

\def\OFourA{O4a\xspace}

\def\gwh{GW\xspace}
\def\cwh{CW\xspace}

\newcommand{\tcr}{\textcolor{black}}
\newcommand{\tcb}{\textcolor{black}}

\def\gws{GWs\xspace}

\def\uldm{ultralight DM\xspace}
\def\uldmh{ultralight DM\xspace}

\def\dm{DM\xspace}
\def\dmh{DM\xspace}
\def\lvk{LIGO, Virgo and KAGRA\xspace}
\def\ifo{interferometer\xspace}
\def\ifos{interferometers\xspace}

\def\psd{power spectral density\xspace}

\def\nn{\nonumber}

\def\sfdbs{short fast Fourier transform databases\xspace}

\def\bsds{BSDs\xspace}
\def\sfts{SFTs\xspace}

\def\xcorr{\emph{cross-correlation}\xspace}
\def\Xcorr{\emph{Cross-correlation}\xspace}
\def\ep{\emph{BSD-excess-power}\xspace}
\def\Ep{\emph{BSD-excess-power}\xspace}
\def\lpsd{\emph{LPSD}\xspace}

\newcommand{\TFFT}{T_\text{FFT}}

\newcommand{\Tobs}{T_\text{obs}}

\newcommand{\be}{\begin{equation}}
\newcommand{\ee}{\end{equation}}
\newcommand{\rhoDM}{\rho_{\text{DM}}}
\newcommand{\mdm}{m_{\rm DM}}
\newcommand{\Tcoh}{T_{\text{coh}}}

\newcommand{\homin}{h^{95\%}_{\rm 0,min}}

\newcommand{\ev}{\,\mathrm{eV}}

\newcommand{\hz}{\,\mathrm{Hz}\,}

\newcommand{\mpl}{m_{\text{Pl}}}

\newtoggle{fullauthorlist}
\toggletrue{fullauthorlist}
\newtoggle{endauthorlist}
\toggletrue{endauthorlist}

\begin{document}
\title{Direct multi-model dark-matter search with gravitational-wave interferometers \\ using data from the first part of the fourth LIGO-Virgo-KAGRA observing run}
\date{\today}

\iftoggle{endauthorlist}{
  %
  %
  \let\mymaketitle\maketitle
  \let\myauthor\author
  \let\myaffiliation\affiliation
  \author{The LIGO Scientific Collaboration}
  \author{The Virgo Collaboration}
  \author{The KAGRA Collaboration}
  \email{full author list given at the end of the article.}
\noaffiliation
}{
\iftoggle{fullauthorlist}{
}
{
  \author{The LIGO Scientific Collaboration}
  \affiliation{LSC}
  \author{The Virgo Collaboration}
  \affiliation{Virgo}
  \author{The KAGRA Collaboration}
  \affiliation{KAGRA}
  
}
}



\graphicspath{{./figures/}}

\begin{abstract}
We present a search of data from the first part of the fourth observing run of LIGO--Virgo--KAGRA for three kinds of dark matter -- dilatons (spin-0), dark photons (spin-1) and tensor bosons (spin-2) -- using three independent methods, \tcb{each extended to search for all three candidates}. Each dark-matter candidate could interact with different standard-model particles in the instruments, causing unique differential strains on the interferometers. While we do not find any evidence for a signal, we place the most stringent upper limits to-date on \tcb{the couplings of each dark-matter model to the interferometers}. For scalars with masses between $[4\times 10^{-14},1.5\times 10^{-13}]$ eV that couple to photons or electrons, our constraints \tcb{on the scalar-to-electron or photon coupling} improve upon those from the third observing run by one order of magnitude, with the tightest limit of $\sim 10^{-20}\,\text{GeV}^{-1}$ at a mass of $\sim2\times 10^{-13}\text{ eV}$. For vectors with masses between $[7\times 10^{-13},8.47\times 10^{-12}]$ eV that couple to baryons, our constraints \tcb{on the dark-photon/baryon coupling} supersede those from MICROSCOPE and Eöt-Wash by one to two orders of magnitude, reaching a minimum of $\sim 5\times 10^{-24}$ at a mass of $\sim 10^{-12}$ eV. For tensors with masses of $[4\times 10^{-14},8.47\times 10^{-12}]$ eV (the full mass range analyzed) that couple via a Yukawa interaction, our constraints\tcb{, which are the first obtained from gravitational-wave interferometers}, surpass those from fifth-force experiments by four to five orders of magnitude, achieving a limit as low as $\sim 8\times 10^{-9}$ at $\sim2\times 10^{-13}$ eV. \tcb{Our results show that each method has substantially different sensitivity to each dark-matter candidate, highlighting the importance of a multi-method approach to search for new physics. Moreover, they demonstrate that gravitational-wave interferometers can simultaneously probe multiple ultralight dark-matter models and place unprecedented constraints on tensor fields arising in a broad range of modified-gravity theories. }
\end{abstract}
        
\maketitle

\section{Introduction}\label{sec:intro}

The existence of dark matter (\dm) is well established, yet its fundamental nature remains one of the most pressing mysteries in modern physics \cite{Bertone:2016nfn}. Across a vast range of possible masses and interaction strengths, \dm could take many forms -- from particles with masses of $\mathcal{O}(10^{-22})$ eV, whose wave-like behavior spans galactic scales \cite{Ferreira:2020fam}, to primordial black holes with masses comparable to stars \cite{Carr:2016drx,Carr:2019kxo,Green:2020jor}. Among these possibilities, \uldm with masses $\mdm\ll 1\ev$ presents a particularly compelling scenario. Not only do high-energy theories naturally predict such particles, but their extraordinarily low mass also implies a quantum-mechanical coherence over macroscopic distances, effectively turning them into oscillating background fields. If these fields interact weakly with standard-model particles, they could produce distinctive, nearly monochromatic signals detectable by precision experiments \cite{Pierce:2018xmy,Grote:2019uvn}.

Gravitational-wave (GW) detectors, such as \lvk \cite{2015CQGra..32g4001L,2015CQGra..32b4001A,KAGRA:2020tym}, have already reshaped astrophysics by measuring the signals from merging black holes and neutron stars \cite{Abbott:2016blz,gw170817FIRST,LIGOScientific:2021djp}. But their extraordinary sensitivity also makes them powerful tools to probe fundamental physics beyond \gws \cite{Arvanitaki:2014wva,Bertone:2019irm,Nguyen:2024fpq,Miller:2025yyx}. The same precision that captures the faint signals from GWs could reveal subtle interactions between DM and ordinary matter, \tcb{and probe small modifications to gravity}, which places these detectors at the forefront of searches for new physics. Along these lines, stringent constraints have been set on dilatons and dark photons from searches on past \gwh data \cite{Guo:2019ker,LIGOScientific:2021odm,Vermeulen:2021epa,Aiello:2021wlp,Fukusumi:2023kqd,Gottel:2024cfj}.

\tcb{The \uldm candidates considered here occupy a portion of the parameter space that is largely inaccessible to previous experiments. Indirect-\dm searches via gamma-ray emission, e.g. VERITAS or Fermi, target WIMPs in the GeV-TeV mass range \cite{PerezdelosHeros:2020qyt}, while experiments such as NuSTAR constrain sterile-neutrino \dm at masses of $\mathcal{O}($keV)  \cite{Ng:2019gch}. Additionally, ADMX \cite{Du:2018uak} and ALPS \cite{Hartman:2020bgj} have searched for axion \uldm with masses of $\mathcal{O}(\mu$eV) and $\mathcal{O}$(meV), respectively, which also lie outside the mass range that we study.}

\tcb{Likewise, existing \gwh searches with \lvk are not sensitive to the \uldm models considered here, despite targeting signals with similar morphologies. These searches are designed to find Doppler-shifted, spinning-down signals from rapidly rotating neutron stars or boson clouds around spinning black holes \cite{Riles:2022wwz}, which do not match the signatures of \uldmh signals. Additionally, narrow-band stochastic \gwh background searches use a frequency-dependent overlap reduction function (ORF) that differs from that of \uldmh interactions\footnote{The ORF quantifies the way that \gws or \dm couple to the \ifos and is described in detail in the Supplemental Material.} \cite{LIGOScientific:2025bgj,LIGOScientific:2025bkz}, and much shorter analysis coherence times that are not optimized for \uldm. Moreover, these searches only consider real, positive signal-to-noise ratios (SNRs) as being astrophysical, whereas both dark photons and tensor \dmh would generate negative SNRs because of their ORFs \cite{Manita:2022tkl}, causing standard analyses to miss them \cite{LIGOScientific:2025bgj,LIGOScientific:2025bkz}.}

In contrast to previous works, we explore here how \gwh detectors can be used to search for \uldm in various forms simultaneously: scalar fields (such as dilatons, which may induce oscillations in fundamental constants) \cite{Stadnik2015a,Stadnik2015b,Stadnik2016, Grote:2019uvn}, vector fields (like dark photons, exerting weak oscillating forces on matter) \cite{Pierce:2018xmy}, and massive tensor fields (closely resembling \gws but with a nonzero mass) \cite{Aoki:2016zgp,Marzola:2017lbt,Armaleo:2020efr,Manita:2023mnc}, \tcb{which appear in a broad range of modified gravity theories}. Each of these candidates leaves a unique imprint on \gwh detectors, offering a new way to constrain -- or perhaps discover -- some of the most elusive \dm candidates predicted by theory.

\section{Dark-matter interaction models} \label{sec:model}

Cold, \uldm consists of an enormous number of extremely light particles, which motivates its treatment as a classical oscillating field. The frequency of this field is fixed by the \uldmh mass \tcb{regardless of the type of \dm considered}: $f_0 =\mdm c^2/h$; however, there is a small, stochastic frequency change caused by the Maxwell-Boltzmann velocity distribution of the \dmh particles: $\Delta f/f\sim v_0^2/c^2\sim 10^{-6}$, where $v_0$ is the virial velocity of \dm around the center of the galaxy. \tcb{This frequency spread cannot be captured by standard deterministic methods that look for quasi-monochromatic signals without incurring a sensitivity loss}. Moreover, the signal has a finite coherence time \tcb{that scales inversely with its mass}, of $\mathcal{O}(10^4)$ s at $\mdm\sim 10^{-12}$ eV \cite{Miller:2025yyx}, in which the signal can be treated as purely monochromatic, but spreads over $\Delta f$ if observing for longer than the coherence time. Likewise, the signal has a finite coherence length, which greatly exceeds the separation of earth-based \ifos, \tcb{implying correlated signals.} 

\tcb{Regardless of the \uldm model, the signal appears in the detector as a persistent, quasi-monochromatic signal with a finite coherence time, and a narrow-band signal with stochastic frequency fluctuations over longer timescales. Each model described below \tcr{for spin-0, 1 and 2} differs in the mechanism by which \dm couples to the \ifos and in the induced strain amplitude, not in the phase evolution.} \tcr{While models for higher-spin \dm exist \cite{Criado:2020jkp,Alexander:2020gmv,Jain:2021pnk}, the ways in which they interact with \gwh \ifos are model-dependent. After specifying the couplings, we can generalize the analyses presented here to constrain them as well.}

\subsection{Scalar, dilaton dark matter}

Models of \uldm predict that it could have originated through a vacuum misalignment mechanism in the early universe, and would manifest today as a coherently oscillating scalar field~\cite{Stadnik2015b}. The coupling of such \dm would effectively induce an oscillation of the values of fundamental constants, namely the electron rest mass and the fine structure constant, thus leading to the expansion and contraction of solids. When a field gradient exists, objects can also be accelerated. Many ideas have been proposed to detect such effects~\cite{Preskill:1982cy,Abbott:1982af,Dine:1982ah,Cho:2007cy}, with the current strongest direct constraints coming from experiments looking for the oscillation of fundamental constants using atomic clocks or spectroscopy~\cite{Arvanitaki:2014faa,Aharony:2019iad,Savalle:2020vgz,Kennedy:2020bac,Antypas:2019qji,Antypas:2020rtg,Tretiak:2022ndx,Oswald:2021vtc,Campbell:2020fvq,BACON:2020ubh,Zhang:2022ewz,Sherrill:2023zah}. More recently, \tcb{several experimental mechanisms, in particular the way in which dilatons could couple to the beamsplitter}, were discovered to detect those same oscillations using \gwh interferometers~\cite{Stadnik2015a,Stadnik2015b,Stadnik2016,Grote:2019uvn,PhysRevD.100.123512,Morisaki:2020gui}. 

 Considering a simple linear expansion to the standard-model Lagrangian $\mathcal{L}$, we write the scalar field $\phi$ as~\cite{Vermeulen:2021epa}:
\begin{equation}\label{eq:L_int}
    \mathcal{L} \supset \frac{\phi}{\Lambda_\gamma} \frac{F_{\mu\nu}F^{\mu\nu}}{4}  - \frac{\phi}{\Lambda_{\rm e}} m_{\rm e} \bar{\psi}_{\rm e} \psi_{\rm e},
\end{equation}
where $F_{\mu\nu}$ is the electromagnetic field tensor, $\psi_{\rm e}$ and $\bar{\psi}_e$ are the standard-model electron field and its Dirac conjugate, respectively, $m_{\rm e}$ is the electron rest mass, and $\Lambda_\gamma$ and $\Lambda_{\rm e}$ denote the scalar \dmh coupling constants to the photon and electron, respectively.
When $\Lambda_i^{-1}$ is non-zero, the \dmh field induces changes in $m_e$ and the fine structure constant through $F_{\mu\nu}$. In turn, this modulates the sizes and refractive indices of solids.

In LIGO, by considering only oscillations in the size of the beamsplitter and mirrors, we can relate the measured strain $h(f_0)$ to $\Lambda_i$ as~\cite{Gottel:2024cfj}:


\begin{align} \label{eq:LPSD:DM_Lambda}
    \left(\frac{1}{\Lambda_\gamma} + \frac{1}{\Lambda_{\rm e}}\right)  &\approx \frac{\tcb{\mdm}c}{\tcb{\hbar}\,\sqrt{2\,\rhoDM}}h(f_0)A_{\rm cal}(f_0) \nn  \\
   &\approx 10^{-18}\,\SI{}{\per\giga\electronvolt}\left(\frac{f_0}{\SI{100}{\hertz}}\right)\left(\frac{h_0(f_0)A_{\rm cal}(f_0)}{10^{-17}\,\SI{}{\per\second}}\right)
\end{align}


where $\rhoDM=\SI{0.4}{\giga\electronvolt\per\cubic\centi\meter}$ is the local \dmh energy density from the standard galactic \dmh halo model \cite{Read:2014qva}.
$A_{\rm cal}$ encodes the \ifo response to scalar \dm \cite{Gottel:2024cfj}, and is obtained through optical simulations that account for finite light travel time effects~\cite{Morisaki:2020gui} (see the Supplemental Material). 
\tcr{Note here we make the minimal assumption about the scalar potential by keeping only the quadratic mass term. This type
of potential can naturally arise when the presence of an
underlying approximate symmetry leads to strongly suppressed
self-interactions, such as the approximate shift symmetry in
relaxion models~\cite{Graham:2015cka}. In such models, the quartic and higher-order terms are small and do not induce a significant change in the scalar mass near the Earth due to the scalar field's linear couplings to matter ~\cite{Khoury:2003rn,Khoury:2003aq,Olive:2007aj,Brax:2010gi,Burrage:2017qrf}.}


\subsection{Vector dark photon dark matter}

Spin-1 dark photons\footnote{\tcb{These U(1)$_B$ or U(1)$_{B-L}$ dark photons couple to baryon or baryon-lepton number and are different than those that kinetically mix with the ordinary photon \cite{Holdom:1985ag}; thus, we do not consider the charge distributions of the materials, nor screening effects.}}
could completely explain the relic abundance of \dm \cite{Agrawal:2018vin}. These particles could arise from the misalignment mechanism \cite{nelson2011dark,arias2012wispy,graham2016vector}, parametric resonance or the tachyonic instability of a scalar field \cite{agrawal2020relic,Co:2018lka,Bastero-Gil:2018uel,Dror:2018pdh}, or from cosmic string network decays \cite{long2019dark}. The observable effect would result from a coupling of dark photons to standard-model particles -- either baryons or baryon minus leptons. In particular, this interaction would cause ``dark'' forces on the mirrors that would oscillate at a frequency fixed by dark photon \cite{Pierce:2018xmy,Miller:2020vsl}. 

The Lagrangian $\mathcal{L}$ that characterizes the dark photon coupling to a number current density $J^\mu$ of baryons or baryons minus leptons is:

\begin{equation}
    \mathcal{L} = -\frac{1}{4\mu_0} F'^{\mu \nu} F_{'\mu \nu} + \frac{1}{2\mu_0} \left(\frac{\mdm c }{\hbar}\right)^2 A^\mu A_\mu - \epsilon e J^\mu A_\mu, \label{eqn:dplagrangian} \\
\end{equation}
where $F_{'\mu \nu}$ is the dark-photon electromagnetic field tensor, $\mu_0$ is the magnetic permeability in vacuum, $e$ is the electron charge, $A_\mu$ is the dark four-vector potential, and $\epsilon$ is the strength of the particle/dark photon coupling normalized by the electromagnetic coupling constant.

The strain on the \ifos caused by a dark photon \dmh signal is \cite{Pierce:2018xmy,Morisaki:2020gui}:

\begin{align}
\sqrt{\langle h^2_{\rm D}\rangle}&=
\tcb{\frac{1}{3\sqrt{2}\pi}}
\frac{Q}{M}\frac{\hbar e}{c^4\sqrt{\epsilon_0}} \sqrt{2\rhoDM}
v_0
\frac{\epsilon}{f_0}, \nonumber \\
&\simeq 6.56\times 10^{-26}\left(\frac{\epsilon}{10^{-22}}\right)\left(\frac{100 \text{ Hz}}{f_0}\right),
\label{eqn:dph0}
\end{align}
where $\epsilon_0$ is the permittivity of free space, and ${Q}/{M}$ is the \tcb{dark} charge-to-mass ratio of the LIGO mirrors \tcb{under the new $U(1)_B$ current.} 
We take ${Q}/{M}=m_p^{-1}$ \tcb{for $U(1)_{\rm B}$ because the mass of the mirror is dominated by protons and neutrons\footnote{For $U(1)_{\rm B-L}$ currents, ${Q}/{M}=1/(2m_p)$.}}, where
$m_p$ is the proton rest mass \cite{ParticleDataGroup:2024cfk}.

A second \tcb{effective arm-length modulation} results because light takes a finite amount of time to travel between the input and end mirrors in the \ifo, during which the mirrors have moved in response to \dm \cite{Morisaki:2020gui}.

\begin{align}
\sqrt{\langle h_{\rm C}^2\rangle}&=\frac{\sqrt 3}{ 2} \sqrt{\langle h_{\rm D}^2\rangle}\frac{2\pi f_0 L}{v_0}, \nonumber \\
&\simeq 6.58\times 10^{-25}\left(\frac{\epsilon}{10^{-22}}\right),
\label{eqn:dpfiniteh0}
\end{align}
where $L=4$ km is the arm-length of the LIGO \ifos.
The total strain \tcb{from both physical effects always adds as}: $\langle h_{{\rm total}}^2\rangle = \langle h_D^2\rangle+\langle h_C^2\rangle$. The averages are taken over polarization direction and detector geometry.

\subsection{Tensor dark matter}

\tcb{
A novel spin-2 graviton arising in ghost-free massive gravity theories can simultaneously modify gravity and constitute \dm~\cite{Dubovsky:2004ud,Babichev:2016hir,Hassan:2011zd,DeFelice:2020ecp}.
Scenarios in which massive gravitons comprise all of \dm~have been widely explored~\cite{Manita:2022tkl,Jain:2021pnk,Marzola:2017lbt,Aoki:2016zgp,Aoki:2019snr,Babichev:2016bxi,Kolb:2023dzp,Aoki:2017cnz,Babichev:2016hir},
along with the associated phenomenology and detection methodologies in ground-based or space-based laser interferometers~~\cite{Armaleo:2020efr,Manita:2022tkl,Manita:2023mnc,Zhang:2025fck},
atom interferometers~\cite{Blas:2024kps}, and
pulsar timing arrays~\cite{Sun:2021yra,Wu:2023dnp,Armaleo:2020yml,Xia:2023hov,Cai:2024thd},
and through black-hole superradiant instabilities~\cite{Dias:2023ynv}.
In this work,
we consider spin-2 tensor \dm arising from the Hassan-Rosen bigravity~\cite{Hassan:2011zd}, consisting of one massless and one massive graviton,
where the massive eigenstate with five degrees of freedom could explain \dm~\cite{Marzola:2017lbt}.
}
%

The Fierz-Pauli Lagrangian density \cite{Armaleo:2020yml} describes this tensor field $\chi_{\mu\nu}$:

\begin{align}
\mathcal{L}&=
-\frac{1}{4}\chi^{\mu\nu} \mathcal{E}_{\mu\nu,\alpha\beta}\chi^{\alpha\beta}
-\frac{1}{8}\left(\frac{\mdm c}{\hbar}\right)^2\left( \chi_{\mu\nu}\chi^{\mu\nu}-\chi^2\right) \nn \\ 
&\tcb{-\frac{\alpha}{2c\mpl }\sqrt{\frac{\hbar}{c}}\chi_{\mu\nu}T^{\mu\nu}}
,
\label{eqn:FP_quadratic}
\end{align}
where $\mpl$ is the reduced Planck mass,
\( \chi \equiv \chi^\mu_\mu \), \( \mathcal{E}_{\mu\nu,\alpha\beta} \) is a second-order derivative operator defined in Eq.~(2.13) of \cite{Aoki:2016zgp}, \tcb{and $\alpha$ is the Yukawa coupling to the energy-momentum tensor $T_{\mu\nu}$. This kind of coupling is generic to ultralight tensor \dmh models \cite{Aoki:2016zgp}.}
%

The strain on \ifos arises analogously to that from \gws, a stretching of space-time in the presence of the field.
%
%
Specifically,
\tcb{ignoring the finite-light-travel-time effect, the differential strain after averaging over five polarization components can be expressed as }\cite{Aoki:2016zgp,Manita:2023mnc}\footnote{Note that Refs. \cite{Aoki:2016zgp,Manita:2023mnc} and Refs. \cite{Marzola:2017lbt,Armaleo:2020efr} differ by a factor of 2 in the strain induced by tensor \dm, which results from a factor of 8 difference in the Lagrangians. Here, we use the prescriptions in \cite{Aoki:2016zgp,Manita:2023mnc}, which leads to a factor of two stronger upper limits than would be obtained using the formalism of \cite{Marzola:2017lbt,Armaleo:2020efr}.}:

\begin{align}
    h(f_0)&=\frac{2\alpha
    }{ 2\pi f_0 \mpl}
    \sqrt{\frac{\rhoDM \hbar c}{\tcb{5}}} \nn
    \\
    &
    \simeq 
    \tcb{7.78\times 10^{-26}}
    \left(\frac{\alpha}{5\times 10^{-8}}\right)\left(\frac{100\hz}{f_0}\right).
     \label{eq:signal}
\end{align}
%

\section{Data}\label{sec:data}

We analyze data from the first part of the fourth observing run (\OFourA) of the LIGO Livingston (L1) and Hanford (H1) detectors. The sensitivity of LIGO in \OFourA improved significantly across all frequencies, particularly above 400 Hz \cite{LIGOO4Detector:2023wmz,membersoftheLIGOScientific:2024elc,Capote:2024rmo}. The strain data were collected between May 24, 2023, at 15:00 UTC, and January 16, 2024, at 16:00 UTC, with L1 and H1 operating 69.0\% and 67.5\% of the time, respectively \cite{GWOSC:O4a}. Virgo did not take data during \OFourA, and KAGRA observed for one month. The data are calibrated by directing a laser onto the test masses and comparing the measured response to the expected one \cite{Karki:2016pht,Viets:2017yvy}. Additionally, the data have been cleaned to remove ``glitches'', short-duration disturbances \cite{LIGO:2024kkz}, and analyzed only if they are in ``science mode'' \cite{GoetzRiles:T2400058:v1}.
At worst, amplitude and phase uncertainties at $1\sigma$ are 10\% and 10 degrees, respectively, in the frequency range analyzed by all methods: [10, 2000] Hz.

In this analysis, we use three independent pipelines, described in \cref{sec:method}, to search for direct \dmh interactions with \gwh \ifos. Each one employs data in different formats.
The first, called the \Ep method, uses Band Sampled Data (BSD) structures \cite{Piccinni:2018akm} as inputs to this search, which are derived from \sfdbs \cite{Astone:2005fj}. \bsds consist of complex-valued time-series representations of the strain data in 10-Hz bands, generated monthly from the \verb|GDS-CALIB_STRAIN_CLEAN_AR| channel. 
The other two methods, \xcorr and Logarithmic Power Spectral Density (\lpsd), use \emph{gated}\footnote{Gating refers to removing the cumulative effect of many loud, transient glitches that can collectively raise the noise floor and inhibit the recovery of \cwh signals.} data from the channel \verb|GDS-CALIB_STRAIN_CLEAN_GATED_G02|, in which spurious glitches have been removed \cite{DavisEtAl2024_selfgating_O4a,DCHgate2025}.
\Xcorr employs \sfts of length $T_{\rm SFT}=1800$, totalling 5696 coincident H1--L1 \sfts (\SI{2848}{\hour}).
In contrast,  \lpsd requires continuous segments of at least $10^5$ seconds, resulting in 30 coincident segments (\SI{1157}{\hour}).

\section{Search methods} \label{sec:method}

We present three independent methods that searched for \uldm coupling to the \ifos. All methods are sensitive to each kind of \dm presented in \cref{sec:model} \tcb{because each only looks for a monochromatic signal, which is a shared feature of \uldmh signals coupling to \gwh \ifos.} \tcr{The first, \xcorr, relies more heavily on the detector response to each kind of \dm and is therefore inherently model-dependent. In contrast, \ep and \lpsd are generically sensitive to narrow-band \uldm, allowing null results to be more easily reinterpreted as constraints on different \dmh couplings. For \xcorr, extending the search to different kinds of \dm required model-specific adaptations to the search, including the derivation of new ORFs. In contrast, for \ep and \lpsd, the search strategies are unchanged relative to previous works. For all three methods, the resulting strain upper limits are translated into constraints on all three \dmh couplings for the first time.
}

\subsection{Cross-Correlation} \label{subsec:cross-correlation}
\Xcorr is a natural technique to search for \uldm because the signal is expected to correlated amongst multiple detectors \cite{Pierce:2018xmy}, and was used in O1 and O3 DM searches \cite{Guo:2019ker,LIGOScientific:2021odm}. By cross correlating the data in each frequency bin $j$, we can calculate the SNR as
\begin{equation}
\text{SNR}_{j} = \frac{S_{j}}{\sigma_{j}} .
\label{eq:ccsnr}
\end{equation}
$S_j$ is the signal strength constructed by cross-correlating FFT datasets 
$z_{1,ij}$ and $z_{2,ij}$ of detectors 1 and 2 for segment $i$, 
and then averaging over the segments:
\begin{eqnarray}
S_j = \frac{1}{N_{\text{SFT}}} \sum_{i=1}^{N_{\text{SFT}}} \frac{z_{1,ij} z_{2,ij}^{\ast}}{P_{1,ij} P_{2,ij}},
\end{eqnarray}
where $N_{\text{SFT}}$ is the number of SFTs used, $P_{1(2),ij}$ is the noise power estimated from the neighboring 50 frequency bins\footnote{Fifty bins ensures that the signal does not pollute the background estimation and the noise properties do not differ by too much across the frequency range of the estimation.} using a running 
median, and is related to the one-sided power spectral density (PSD) by 
$\text{PSD}_{1(2),ij} = {2P_{1(2),ij}}/{T_{\text{SFT}}}$. When a signal is not present, the expectation value of
$S_j$ is zero, and the variance for its real and imaginary parts is 
\begin{eqnarray}
\sigma_j^2 = \frac{1}{N_{\text{SFT}}} \left\langle \frac{1}{2P_{1,ij} P_{2,ij}} \right\rangle_{N_{\text{SFT}}}.
\end{eqnarray}
For scalar \dm, the SNR is positive, whereas for dark photons and tensor bosons, it is negative because the corresponding ORF is $\sim -0.9$ over the analyzed frequency range. \tcb{These ORFs are derived in the Supplemental Material.}


\subsection{BSD excess power method}\label{subsec:excess}

As done in O3 \cite{LIGOScientific:2021odm}, in each 1-Hz band, \ep changes the fast Fourier transform coherence time $\TFFT$ to match the \dmh signal coherence time $\Tcoh$, ensuring that all power is confined to one frequency bin for the full observing run for each \dm mass \cite{Miller:2020vsl}. We then take Fourier transforms of the strain data using \tcb{$\TFFT\in [10^{2.75},10^{4.5}]$ s at high and low frequencies, respectively \cite{Miller:2020vsl,LIGOScientific:2021odm},} to make time-frequency ``peakmaps'' \cite{Astone:2005fj,Astone:2014esa}. ``Peakmaps'' are collections of ones that contain frequency bins whose powers exceed a threshold in the equalized spectrum, i.e. \tcb{the ratio of the signal power normalized by the noise power, estimated with a running median \cite{Miller:2020vsl}}, and are also local maxima. We then count the number of peaks at each frequency to find bins with large numbers of peaks. By counting the peaks, and not the  power in each frequency bin, we are more robust against non-Gaussian noise artifacts that could contaminate particular frequency bins and bias our detection statistic, the critical ratio $CR$:


\begin{equation}
    CR=\frac{y-\mu}{\sigma}.
\end{equation}
Here, $y$ is the number of peaks at a particular frequency, and $\mu$ and $\sigma$ are the mean and standard deviations of the number counts across all frequencies in the band, respectively. In Gaussian noise, the $CR$ is normally distributed with zero mean and unit variance \cite{Astone:2014esa}, \tcb{and a signal would appear in the positive tail of this distribution.}

We select enough candidates\footnote{The number of candidates selected ranges from $20$ and $285$ at 2000 Hz and 10 Hz, respectively and depends on $\TFFT$.} in each 1-Hz band such that one coincident candidate between two detectors would occur in Gaussian noise. This ensures that we select uniformly over the frequency range.

\subsection{LPSD} \label{subsec:LPSD}
\textit{LPSD} exactly matches the integration time in Fourier transforms to the \dmh coherence time, \tcb{similarly to what \ep does,} but in every \tcb{logarithmically-spaced frequency} bin, \tcb{not every Hertz}. This allows it to achieve optimal sensitivity at each \dmh mass, \tcb{regardless of the type of \dm}. However, the logarithmic frequency spacing poses a challenge because FFT-based algorithms are no longer applicable. With a computational complexity scaling as $\mathcal{O}(N^2)$, where $N$ is the number of data points, this leads to intractable costs. However, we developed a method that leverages symmetries between the time and frequency domains~\cite{Gottel:2025aqm}. This allows us to perform the full logarithmic calculation with a single FFT followed by a heavily zero-suppressed transformation without the need for expensive pre-computations. 

We then follow the approach in~\cite{Gottel:2024cfj,Gottel:2025aqm}, where a background model is constructed by fitting cubic splines in log-log space to the PSD as a function of frequency. Because the PSDs vary by several orders of magnitude, we fit over intervals of $10^4$ frequency bins. Within these intervals, potential \dm signals are identified as positive deviations from the fits by using a profile likelihood test (see Eq. 15 in~\cite{Gottel:2025aqm}), assuming, as validated empirically, that the residuals follow a skew-normal distribution.

\section{Results}\label{sec:results}

\subsection{Cross-Correlation}
We set thresholds for detection of $\text{Re}(\text{SNR}) < -5.8$ \tcb{for dark photons and tensor bosons, and $\text{Re}(\text{SNR}) > 5.8$ for scalar bosons} that correspond to $\sim1\%$ false-alarm probability, but find no candidates. We also checked subthreshold outliers with SNRs in the range $[5.0, 5.8]$ for either the real or imaginary part of the 
SNR. Of ten such outliers, nine were due to loud instrumental artifacts, and the remaining one was consistent with Gaussian noise. 


In the absence of a detection, and after excluding frequency bins lying within $\sim 0.056$ Hz of known spectral artifacts, the measured SNR is converted into upper limits on each of the aforementioned \dmh couplings using the Feldman-Cousins formalism \cite{Feldman:1997qc}. This conversion incorporates both an efficiency factor obtained from simulations to correct for the power loss due to the frequency binning choice and random frequency-bin boundary, and the appropriate ORF.
These limits are set bin-by-bin and thus exhibit larger fluctuations than limits calculated in 1-Hz bands by the other methods.


\subsection{BSD Excess power method}

Our search returned 6847 coincident candidates present in the same frequency bin in both L1 and H1 and that did not overlap with known noise disturbances \cite{LIGO:2024kkz}. However, after requiring that the average $CR$ be greater than five (corresponding to $5\sigma$ significance in Gaussian noise), only ten candidates remained. To determine whether these candidates resulted from \dm, we performed three follow-up procedures: (1) we varied the observation time $\Tobs$ at the chosen $\TFFT$ to see if there was a steady increase in the $CR\propto \Tobs^{1/2}$, (2) we increased $\TFFT>\Tcoh$ to see if there was a \emph{decrease} in the $CR$ \cite{Miller:2020vsl,Vermeulen:2021epa}, and (3) we correlated the spectra from L1 and H1 with an expected \dmh waveform using the Wiener filter \cite{wiener1964extrapolation} and computed the residual between them \cite{Miller:2022wxu}. No candidate survived all three follow-ups.

In the absence of candidates, we set Feldman-Cousins \cite{Feldman:1997qc} upper limits  on the minimum detectable strain amplitude at 95\% confidence $\homin$ induced by a generic \dmh signal using Eq. 75 of \cite{Miller:2025yyx}, which depends on the detector \psd, the maximum of the coincident candidates' $CRs$, $\TFFT$, and $\Tobs$. Up to an $\mathcal{O}(1)$ factor that depends on the \dm polarization and the geometry of LIGO, these limits can then be mapped to \emph{any} kind of \dmh interaction that would induce a differential strain on the detector.

\subsection{LPSD} \label{sec:results:LPSD}

We obtain candidates by performing a profile likelihood ratio test in each bin, as outlined in~\cite{Gottel:2025aqm}, and require a $5\sigma$ significance corrected for the \textit{look-elsewhere effect}
in rejecting the no-DM hypothesis. After clustering neighboring bins, requiring significance and compatibility in both detectors, we find only one candidate which was found to originate from a non-optimal setting in the background fit. 

The resulting 95\% confidence-level upper limits on strain amplitude can then be derived. The use of the gating procedure  led to a factor of $\sim 3$ improvement in the LPSD upper limits with respect to those in O3 \cite{Gottel:2024cfj}.
Injection tests were performed that confirmed the upper limits at the expected level, but found a systematic 13\% overestimate of the limits. This is small compared to the 30\% statistical uncertainty.

\subsection{Upper limits}

\begin{figure*}[ht!]
     \begin{center}
        \subfigure[\,Dilatons]{%
            \label{fig:dilaton-lims}

        \includegraphics[width=0.31\linewidth]{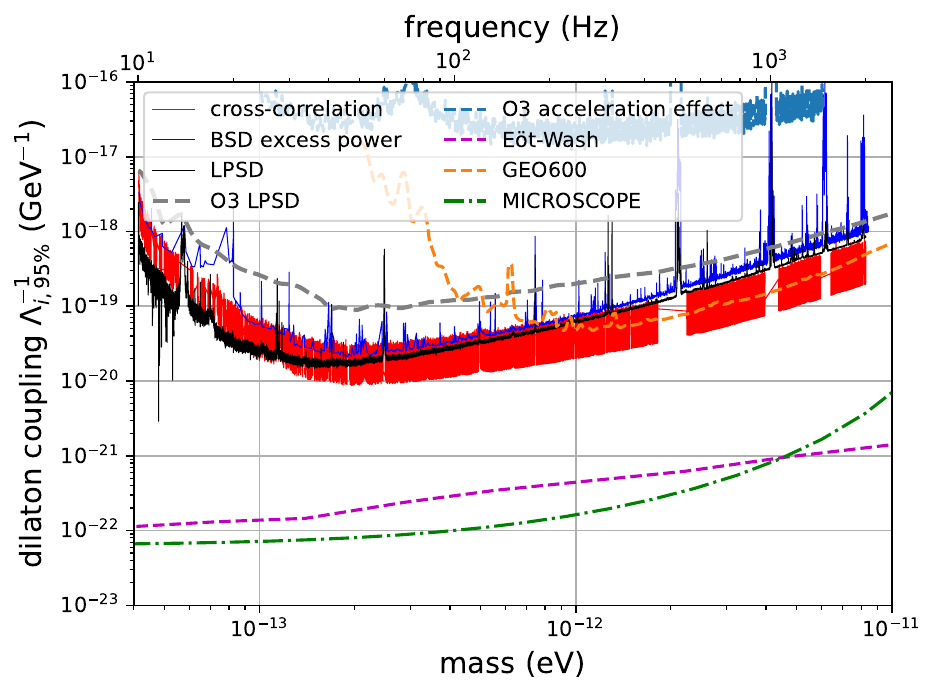}
        }%
        \subfigure[\, Dark photons]{%
           \label{fig:dp-lims}
           \includegraphics[width=0.31\linewidth]{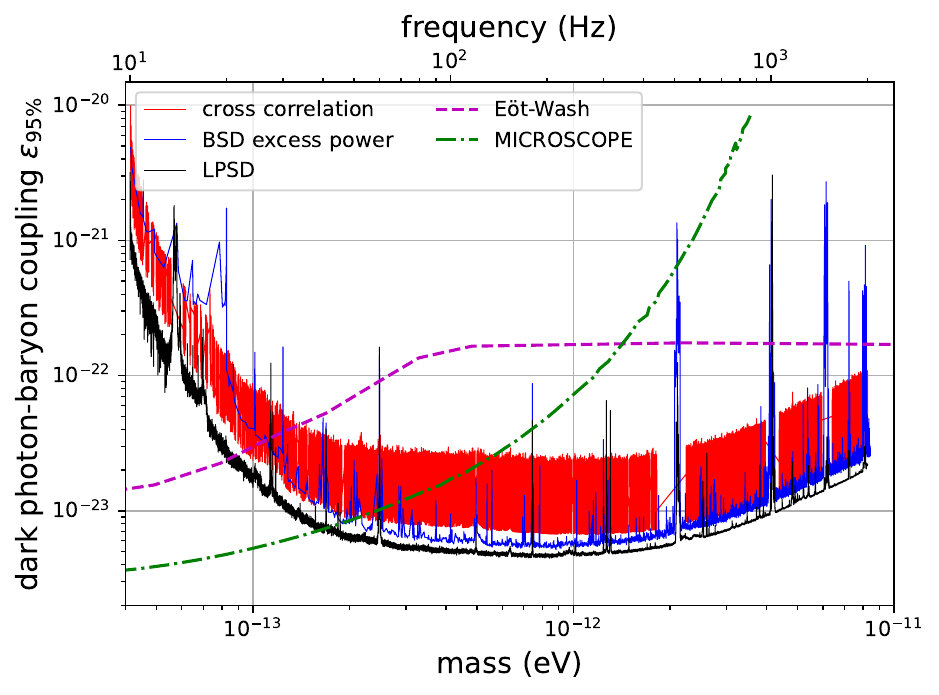}
        }
        \subfigure[\,Tensor bosons]{%
           \label{fig:tensor-dm-lims}
           \includegraphics[width=0.31\linewidth]{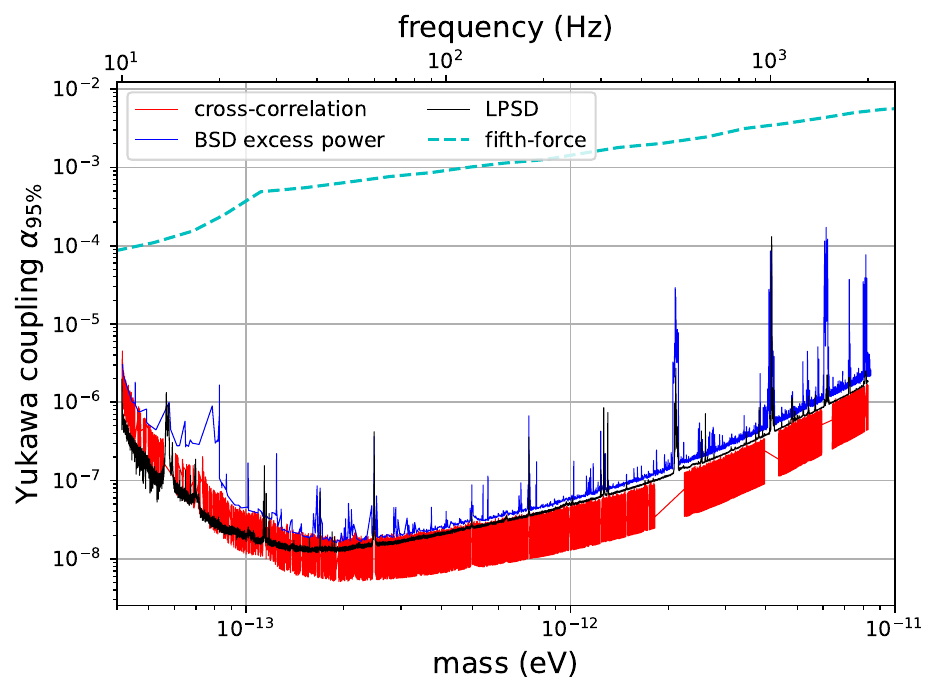}
        }\\ 
    \end{center}
    \caption[]{\textbf{Upper limits on the coupling strengths of ultralight scalar, vector and tensor bosons to standard-model particles in LIGO}. Solid lines denote the constraints from this work (``cross-correlation,'' ``BSD excess power,'' and ``LPSD''). The dashed and dotted-dashed lines indicate constraints from other experiments (``Eöt-Wash'' and ``MICROSCOPE'') and from previous searches on \gwh data (``GEO600'', ``O3 LPSD''\footnote{We note that the ``O3 LPSD curve'' is multiplied by a factor 2.226 compared to the results in~\cite{Gottel:2024cfj} after we found a conversion error.} and ``O3 acceleration effect''). (a) Our constraints apply to the coupling of dilatons to both photons and to electrons. They improve upon those from previous searches with \gwh \ifos \cite{Vermeulen:2021epa,Fukusumi:2023kqd,Gottel:2024cfj} and beat those from O3 by one order of magnitude.  Constraints from Eöt-Wash \cite{Su:1994gu,Schlamminger:2007ht} and MICROSCOPE \cite{Touboul:2012ui,Berge:2017ovy} correspond to the coupling of the dilaton to the electron. (b) The constraints on vector dark photon \dm improve upon existing limits by a couple of order of magnitudes in the dark-photon/baryon coupling constant. To obtain limits on dark photon/baryon minus lepton coupling, $U(1)_{\rm B-L}$, these limits should be multiplied by \tcb{approximately} two, \tcb{but depends on the particular mirror composition}. (c) For the first time, we derive constraints on tensor \dm using \gwh \ifos, which are up to five orders of magnitude superior to those obtained indirectly from fifth-force experiments \cite{Sereno:2006mw,Murata:2014nra}. We release the data to make these figures as well \cite{Guo:2025O4aDMData}.
    }%
     \label{fig:upper-limits}
\end{figure*}

 In \cref{fig:upper-limits}, we show upper limits for each of the three \dmh models from each pipeline, as argued in \cite{rellathesis,Miller:2025yyx}. \tcb{The semi-analytic way that these limits are computed is detailed in \cite{Miller:2025yyx}, and holds even though the data are not Gaussian, as justified by injection studies \cite{Guo:2019ker,Miller:2020vsl,Gottel:2024cfj}}. For all three \dmh models, we see significant improvements in the constraints on the coupling constants with respect to existing ones. For dilatons, \xcorr is consistent with or outperforms \ep and \lpsd because all three methods are equally sensitive to the finite light-travel time effect. For dark photons, \xcorr obtains less stringent constraints than those from \ep and \lpsd because it is much less sensitive to the finite-light travel time effect \cite{Morisaki:2020gui,Manita:2023mnc}. For tensor bosons, the upper limits obtained by our pipelines are consistent and significantly outperform fifth-force experiments by several orders of magnitude. 

We note that the upper limits \tcb{resulting from tests of the equivalence principle} are stronger than those obtained from our search for scalar \dm.
\tcb{In this case, our sensitivity is driven by size modulations of the beamsplitter. In laser \ifos, light travels back and forth $\mathcal{O}(10^2)$ times in the Fabry-Perot cavities per encounter with the beamsplitter, which significantly reduces our sensitivity. Because of the difference in test-mass thicknesses \cite{Gottel:2024cfj}, and the better low-frequency sensitivity and longer observing time, we obtain a substantial improvement below \SI{200}{\hertz} compared to the GEO600~\cite{Vermeulen:2021epa} search, which did not suffer from Fabry-Perot suppression.}


By contrast, the sensitivity to dark photons and tensor bosons depends primarily on the interferometer arm length, just as for \gws: \tcb{longer arms imply that smaller displacements of induced \dmh signals can be measured}. In these cases, the differential strain induced by \dm coupling to the test masses, or spacetime distortions from tensor fields, is amplified by the long arms. In particular, for tensor bosons, a torsion balance responds as a single bulk object \tcb{because the coherence length of the \dmh field is much larger than the size of the experiment \cite{Co:2018lka}, leading to an extremely small strain}. This is analogous to resonant-bar detectors, which are far less sensitive than \ifos \cite{maggiore2008gravitational}. Thus, fifth-force experiments are stronger probes of scalars than of vectors and tensors, while \gwh \ifos are more powerful probes of vectors and tensors than scalars.

\section{Conclusions}\label{sec:conclusions}
We have shown that \gwh \ifos can probe multiple kinds of \dm, \tcb{including tensor bosons arising from a broad range of modified theories of gravity}, using three independent search techniques.
\tcb{In particular, we have demonstrated that each of our methods, previously tailored to search for only one particular \dmh model, can be extended to probe a wide range of \dmh interactions, enabling a systematic comparison of their sensitivities to scalar, vector, and tensor signals. \tcr{\Xcorr is tied explicitly to the signal model through the ORFs and detector response, derived in the Supplemental Material, and can be used to distinguish among \uldmh models \cite{Miller:2022wxu,Manita:2023mnc}. In contrast, \ep and \lpsd are sensitive to generic monochromatic narrow-band signals, independent of the underlying \uldmh model. This allows null results to be reinterpreted as constraints on the three \dmh models considered here, which is done for the first time. Taken together, these three methods combine model-dependent and model-agnostic approaches to search for \uldm.}} 

\tcb{Moreover, we found that each method exhibits substantially different sensitivities to each kind of \dm, highlighting the importance of developing multiple independent search pipelines to probe new physics with \gwh detectors.}
Specifically, for dilatons and dark photons, detector improvements and increasingly sensitive methods allow us to probe progressively weaker couplings. For tensor bosons, \gwh \ifos now surpass fifth-force constraints by up to five orders of magnitude, \tcb{establishing them as competitive probes of spin-2 \dm and modified gravity theories}.

\tcb{
Our chosen tensor-field model serves as a benchmark for a broad range of modified gravity theories, paving the way for our results to be reinterpreted to constrain other tensor fields. We have thus demonstrated that \lvk can act not only as direct \dmh detectors, but also as direct probes of modified theories of gravity in previously unexplored ways. } 


\section*{Acknowledgments}

This material is based upon work supported by NSF's LIGO Laboratory, which is a
major facility fully funded by the National Science Foundation.
The authors also gratefully acknowledge the support of
the Science and Technology Facilities Council (STFC) of the
United Kingdom, the Max-Planck-Society (MPS), and the State of
Niedersachsen/Germany for support of the construction of Advanced LIGO 
and construction and operation of the GEO\,600 detector. 
Additional support for Advanced LIGO was provided by the Australian Research Council.
The authors gratefully acknowledge the Italian Istituto Nazionale di Fisica Nucleare (INFN),  
the French Centre National de la Recherche Scientifique (CNRS) and
the Netherlands Organization for Scientific Research (NWO)
for the construction and operation of the Virgo detector
and the creation and support  of the EGO consortium. 
The authors also gratefully acknowledge research support from these agencies as well as by 
the Council of Scientific and Industrial Research of India, 
the Department of Science and Technology, India,
the Science \& Engineering Research Board (SERB), India,
the Ministry of Human Resource Development, India,
the Spanish Agencia Estatal de Investigaci\'on (AEI),
the Spanish Ministerio de Ciencia, Innovaci\'on y Universidades,
the European Union NextGenerationEU/PRTR (PRTR-C17.I1),
the ICSC - CentroNazionale di Ricerca in High Performance Computing, Big Data
and Quantum Computing, funded by the European Union NextGenerationEU,
the Comunitat Auton\`oma de les Illes Balears through the Conselleria d'Educaci\'o i Universitats,
the Conselleria d'Innovaci\'o, Universitats, Ci\`encia i Societat Digital de la Generalitat Valenciana and
the CERCA Programme Generalitat de Catalunya, Spain,
the Polish National Agency for Academic Exchange,
the National Science Centre of Poland and the European Union - European Regional
Development Fund;
the Foundation for Polish Science (FNP),
the Polish Ministry of Science and Higher Education,
the Swiss National Science Foundation (SNSF),
the Russian Science Foundation,
the European Commission,
the European Social Funds (ESF),
the European Regional Development Funds (ERDF),
the Royal Society, 
the Scottish Funding Council, 
the Scottish Universities Physics Alliance, 
the Hungarian Scientific Research Fund (OTKA),
the French Lyon Institute of Origins (LIO),
the Belgian Fonds de la Recherche Scientifique (FRS-FNRS), 
Actions de Recherche Concert\'ees (ARC) and
Fonds Wetenschappelijk Onderzoek - Vlaanderen (FWO), Belgium,
the Paris \^{I}le-de-France Region, 
the National Research, Development and Innovation Office of Hungary (NKFIH), 
the National Research Foundation of Korea,
the Natural Sciences and Engineering Research Council of Canada (NSERC),
the Canadian Foundation for Innovation (CFI),
the Brazilian Ministry of Science, Technology, and Innovations,
the International Center for Theoretical Physics South American Institute for Fundamental Research (ICTP-SAIFR), 
the Research Grants Council of Hong Kong,
the National Natural Science Foundation of China (NSFC),
the Israel Science Foundation (ISF),
the US-Israel Binational Science Fund (BSF),
the Leverhulme Trust, 
the Research Corporation,
the National Science and Technology Council (NSTC), Taiwan,
the United States Department of Energy,
and
the Kavli Foundation.
The authors gratefully acknowledge the support of the NSF, STFC, INFN and CNRS for provision of computational resources.

This work was supported by MEXT,
the JSPS Leading-edge Research Infrastructure Program,
JSPS Grant-in-Aid for Specially Promoted Research 26000005,
JSPS Grant-in-Aid for Scientific Research on Innovative Areas 2402: 24103006,
24103005, and 2905: JP17H06358, JP17H06361 and JP17H06364,
JSPS Core-to-Core Program A.\ Advanced Research Networks,
JSPS Grants-in-Aid for Scientific Research (S) 17H06133 and 20H05639,
JSPS Grant-in-Aid for Transformative Research Areas (A) 20A203: JP20H05854,
the joint research program of the Institute for Cosmic Ray Research,
University of Tokyo,
the National Research Foundation (NRF),
the Computing Infrastructure Project of the Global Science experimental Data hub
Center (GSDC) at KISTI,
the Korea Astronomy and Space Science Institute (KASI),
the Ministry of Science and ICT (MSIT) in Korea,
Academia Sinica (AS),
the AS Grid Center (ASGC) and the National Science and Technology Council (NSTC)
in Taiwan under grants including the Science Vanguard Research Program,
the Advanced Technology Center (ATC) of NAOJ, the Mechanical Engineering Center of KEK and  the
Vietnam National Foundation for Science and Technology
Development (NAFOSTED) 103.01-2025.147.

J.S. is supported by the Peking University under startup Grant No. 7101302974, the NSFC under Grants No. 12025507, No. 12150015, No.12450006.

Additional acknowledgements for support of individual authors may be found in the following document: \\
\texttt{https://dcc.ligo.org/LIGO-M2300033/public}.
For the purpose of open access, the authors have applied a Creative Commons Attribution (CC BY)
license to any Author Accepted Manuscript version arising.
We request that citations to this article use 'A. G. Abac {\it et al.} (LIGO-Virgo-KAGRA Collaboration), ...' or similar phrasing, depending on journal convention.

\appendix
\section{Transfer functions} \label{app:transferfuncs}

In the dilaton search, we use transfer functions to relate the DM-induced fluctuations of the optics and the theorized coupling constants $\Lambda_i^{-1}$ (see \cref{eq:L_int}). Following the approach in~\cite{Gottel:2024cfj}, we use detailed optical simulations in \texttt{Finesse} to calculate the transfer functions between the interferometer's photodiode output amplitude and DM-induced fluctuations in the beamsplitter and arm test mirrors. The results, see \cref{fig:transferfuncs}, clearly show that in O4a, LLO is the more sensitive detector, with roughly a factor three gain in sensitivity (see \cref{sec:results:LPSD}) between O3 and O4a from the transfer function difference alone. At the same time, we also find that the performance of LHO has diminished over the same time period, though only slightly. In both cases, the differences are due to small changes in mirror thicknesses between the two runs.

\begin{figure}[ht!]
    \centering
    \includegraphics[width=\linewidth]{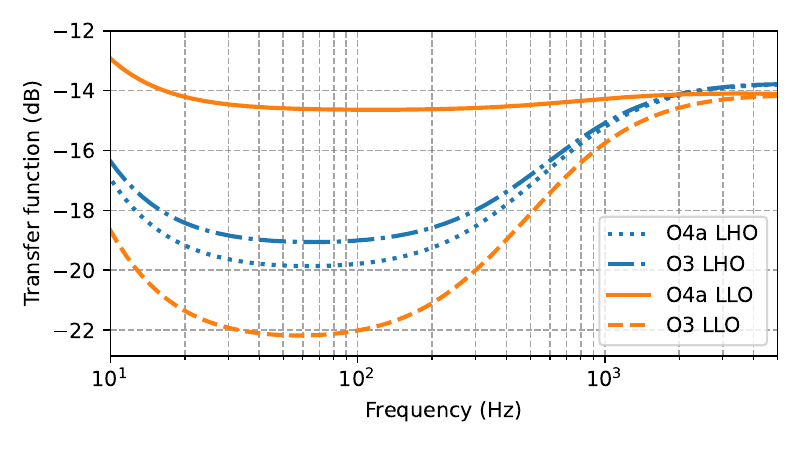}
    \caption{Scalar dark matter transfer functions in dB, considering beamsplitter as well as arm test mirror size oscillations. The dotted lines show the LHO transfer functions, and the others show LLO results. As one can see, while the effect has reduced sligthly in LHO between O3 and O4a, it has increased considerably in LLO, making LLO the more sensitive detector in O4a.}
    \label{fig:transferfuncs}
\end{figure}

{
\section{Overlap reduction functions} \label{app:orf}

In the cross-correlation search, we calculate the SNR by cross-correlating the signal strength between LIGO L1 and H1 detectors. To relate the observed SNR to the sensitivity of the DM coupling constant, one needs the overlap reduction function (ORF). In this section, we derive the ORF that characterizes the cross-correlated DM induced strain signal between LIGO L1 and H1 detectors for the ultralight scalar, vector dark photon, and tensor DM. 

\paragraph{ORF of the scalar DM field.} 
The pattern function of the strain signal induced by the scalar DM field on the size of the beam splitter and the test masses can be found in \cite{Grote:2019uvn,Gottel:2024cfj}, 
\begin{equation}
    F(\hat{\boldsymbol{\Omega}})=1,
\end{equation}
where $\hat{\boldsymbol{\Omega}}$ is the unit vector along the propagation direction of the scalar DM field. The pattern function is independent of $\hat{\boldsymbol{\Omega}}$ and the arm direction of the interferometer because the oscillation of the mirror size is only dependent on the field value of the scalar DM field instead of its gradient.

We define the solid angle average of the pattern function between two detector sites $I,J$,
\begin{equation}
\label{eq:Gamma_IJ}
 \Gamma^{(\rm S)}_{IJ}(f) \equiv  \frac{1}{4\pi}\int d\hat{\boldsymbol{\Omega}}F_I^*(\hat{\boldsymbol{\Omega}})F_J(\hat{\boldsymbol{\Omega}}) e^{2\pi f v \hat{\boldsymbol{\Omega}}\cdot \Delta {\bf x}},
\end{equation}
where $\Delta {\bf x}$ is the displacement vector from the detector site $I$ to the detector site $J$, the subscript labels $I,J$ of the pattern function denote its potential dependence on the detector site position and interferometer arm orientation.
In the frequency of interests in the search, $2\pi f v \hat{\boldsymbol{\Omega}}\cdot \Delta {\bf x}\ll 1$ because the coherence length of the DM field is much longer than the separation of the two detectors, i.e., $|\Delta{\bf x}|/\lambda_{\rm coh}\ll 1$ with $\lambda_{\rm coh}\equiv (2\pi f v)^{-1}$, so the exponential function can be approximated by 1, and thus $\Gamma^{(\rm S)}_{IJ}(f)\simeq \Gamma^{(\rm S)}_{IJ}(0)\equiv \Gamma^{(\rm S)}_{IJ} $. Therefore, the ORF of the scalar DM field is given by
\begin{equation}
\label{eq:orf_s}
    \gamma^{\rm (S)}_{IJ}=\frac{\Gamma^{(\rm S)}_{IJ}}{\sqrt{\Gamma^{(\rm S)}_{II}\Gamma^{(\rm S)}_{JJ}}}=1,
\end{equation}
where the proper normalization is introduced in the denominator.

\paragraph{ORF of the vector dark photon DM field.} The derivation of the ORF of the vector dark photon DM field is similar to the scalar one. The main differences are the polarization sum for the vector case and the pattern functions. There are two signal channels in the vector case, denoted as the common mode and the differential mode signal. The pattern functions are respectively  \cite{Manita:2023mnc},
\begin{eqnarray}
    F^\lambda_{\rm C}(\hat{\boldsymbol{\Omega}})&=&(\hat{\bf m}-\hat{\bf n})\cdot \hat{\bf e}^\lambda(\hat{\boldsymbol{\Omega}}),\nonumber\\
    F^\lambda_{\rm D}(\hat{\boldsymbol{\Omega}})&=&(\hat{\bf m}\otimes\hat{\bf m}-\hat{\bf n}\otimes\hat{\bf n})_{ab} (\hat{\bf e}^\lambda(\hat{\boldsymbol{\Omega}})\otimes\hat{\boldsymbol{\Omega}} )^{ab},~~~~
\end{eqnarray}
where where $\hat{\bf m}, \hat{\bf n}$ are the unit vector along the $X, Y$ arm direction of the detector, $\hat{\bf e}^\lambda(\hat{\boldsymbol{\Omega}})$ is the polarization unit vector of the dark photon field $A^\mu(t,{\bf x})$ with a propagation direction along the unit vector $\hat{\boldsymbol{\Omega}}$, and three polarization modes $\lambda=x,y,z$, e.g., $\hat{\bf e}^z(\hat{\boldsymbol{\Omega}})=\hat{\boldsymbol{\Omega}}$, $\hat{\bf e}^x(\hat{\boldsymbol{\Omega}})$ and $\hat{\bf e}^y(\hat{\boldsymbol{\Omega}})$ are two arbitrary orthogonal unit vectors on the plane perpendicular to $\hat{\boldsymbol{\Omega}}$, $\otimes$ is the tensor product operator, $a,b=1,2,3$, and the Einstein summation convention is adopted here.  

Similar to the scalar case \cref{eq:Gamma_IJ}, one can define the average of the pattern function,
\begin{equation}
    \Gamma^{(\rm V)}_{IJ,s}(f) \equiv \sum_{\lambda} \frac{1}{4\pi}\int d\hat{\boldsymbol{\Omega}}F_{I,s}^{\lambda*}(\hat{\boldsymbol{\Omega}})F^\lambda_{J,s}(\hat{\boldsymbol{\Omega}}) e^{2\pi f v \hat{\boldsymbol{\Omega}}\cdot \Delta {\bf x}},
\end{equation}
where $s={\rm C,D}$ is the signal channel label, and we used the fact that the two channels are orthogonal. In the long coherence length limit, $\Gamma^{(\rm V)}_{IJ,s}(f)\simeq\Gamma^{(\rm V)}_{IJ,s}(0)\equiv \Gamma^{(\rm V)}_{IJ,s}$. The ORF of the vector dark photon DM field is 
\begin{equation}
\label{eq:orf}
    \gamma^{\rm (V)}_{IJ,s}=\frac{\Gamma^{(\rm V)}_{IJ,s}}{\sqrt{\Gamma^{(\rm V)}_{II,s}\Gamma^{(\rm V)}_{JJ,s}}}.
\end{equation}

Given the unit vector of the LIGO L1 and H1 detector arm orientation,
\begin{eqnarray}
    \hat{\bf m}_{\rm H1} = (-0.2239,0.7998,0.5569),\nonumber\\
    \hat{\bf n}_{\rm H1} = (-0.9140,0.0261,-0.4049),\nonumber\\
    \hat{\bf m}_{\rm L1} = (-0.9546,-0.1416,-0.2622),\nonumber\\
    \hat{\bf n}_{\rm L1} = (0.2977,-0.4879,-0.8205),
\end{eqnarray}
one can calculate the ORF explicitly. For the common-mode signal channel in the two LIGO detectors, 
\begin{equation}
    \gamma^{\rm (V)}_{\rm H1L1,C}=-0.02966.
\end{equation}
For the differential-mode signal channel,
\begin{eqnarray}
   \gamma^{\rm (V)}_{\rm H1L1,D} &=&\frac{(\hat{\bf m}\otimes\hat{\bf m}-\hat{\bf n}\otimes\hat{\bf n})|_{{\rm H1},ab}(\hat{\bf m}\otimes\hat{\bf m}-\hat{\bf n}\otimes\hat{\bf n})|_{\rm L1}^{ab}}{\mathcal{N}_2}\nonumber\\
   &=&-0.89,
\end{eqnarray}
where $\mathcal{N}_2=\sqrt{(\hat{\bf m}\otimes\hat{\bf m}-\hat{\bf n}\otimes\hat{\bf n})_{ab}(\hat{\bf m}\otimes\hat{\bf m}-\hat{\bf n}\otimes\hat{\bf n})^{ab}}|_{\rm H1}\times\sqrt{(\hat{\bf m}\otimes\hat{\bf m}-\hat{\bf n}\otimes\hat{\bf n})_{ab}(\hat{\bf m}\otimes\hat{\bf m}-\hat{\bf n}\otimes\hat{\bf n})^{ab}}|_{\rm L1}$ denotes the normalization factor.

\paragraph{ORF of the tensor DM field.}
The pattern functions of the tensor DM field are given in \cite{Manita:2023mnc},
\begin{equation}
    F^\lambda(\hat{\boldsymbol{\Omega}})=D^{ab} \epsilon^\lambda_{ab}(\hat{\boldsymbol{\Omega}}),
\end{equation}
where $D=\frac{1}{2}(\hat{\bf m}\otimes\hat{\bf m}-\hat{\bf n}\otimes\hat{\bf n})$ is the detector tensor, and $\epsilon^\lambda_{ab}(\hat{\boldsymbol{\Omega}})$ is the polarization tensor of the tensor DM field.
Similarly, the ORF of the tensor DM field can be derived by a proper polarization sum over the five polarization modes $\lambda$ of the tensor DM field:
\begin{eqnarray}
    \epsilon^+_{ab}&=&\hat{u}_a \hat{u}_b-\hat{v}_a \hat{v}_b,\nonumber\\
    \epsilon^\times_{ab}&=&\hat{u}_a \hat{v}_b-\hat{v}_a \hat{u}_b,\nonumber\\
    \epsilon^{V_x}_{ab}&=&\hat{u}_a \hat{\Omega}_b-\hat{\Omega}_a \hat{u}_b,\nonumber\\
     \epsilon^{V_y}_{ab}&=&\hat{v}_a \hat{\Omega}_b-\hat{\Omega}_a \hat{v}_b,\nonumber\\
     \epsilon^{\rm s}_{ab}&=&\frac{1}{\sqrt{3}}(\hat{u}_a \hat{u}_b+\hat{v}_a \hat{v}_b-2\hat{\Omega}_a\hat{\Omega}_b),
\end{eqnarray}
where $\hat{\bf u}$ and $\hat{\bf v}$ are two arbitrary orthogonal unit vectors on the plane perpendicular to the propagation direction $\hat{\boldsymbol{\Omega}}$, and $a,b=1,2,3$ label the three components in Cartesian coordinates.

The average of the pattern function is therefore,
\begin{equation}
    \Gamma^{(\rm T)}_{IJ}(f) \equiv \sum_{\lambda} \int\int\frac{d\psi}{2\pi}\frac{d\hat{\boldsymbol{\Omega}}}{4\pi} F_{I}^{\lambda*}(\hat{\boldsymbol{\Omega}})F^\lambda_{J}(\hat{\boldsymbol{\Omega}}) e^{2\pi f v \hat{\boldsymbol{\Omega}}\cdot \Delta {\bf x}},
\end{equation}
where $\psi$ is the polarization angle that defines the reference of the polarization vector $\hat{\bf u}$ and $\hat{\bf v}$. In the long coherence length limit, $\Gamma^{(\rm T)}_{IJ}(f) \simeq \Gamma^{(\rm T)}_{IJ}(0)\equiv \Gamma^{(\rm T)}_{IJ} $. The ORF of the tensor DM field is
\begin{equation}
    \gamma^{\rm (T)}_{IJ}=\frac{\Gamma^{(\rm T)}_{IJ}}{\sqrt{\Gamma^{(\rm T)}_{II}\Gamma^{(\rm T)}_{JJ}}}.
\end{equation}
Substitute the LIGO detector tensor into the above equation, one can find that
\begin{equation}
    \gamma^{\rm (T)}_{\rm H1L1}=-0.89.
\end{equation}
}

\bibliographystyle{apsrev4-1}
\bibliography{references}

@article{Graham:2015cka,
    author = "Graham, Peter W. and Kaplan, David E. and Rajendran, Surjeet",
    title = "{Cosmological Relaxation of the Electroweak Scale}",
    eprint = "1504.07551",
    archivePrefix = "arXiv",
    primaryClass = "hep-ph",
    doi = "10.1103/PhysRevLett.115.221801",
    journal = "Phys. Rev. Lett.",
    volume = "115",
    number = "22",
    pages = "221801",
    year = "2015"
}

@article{ParticleDataGroup:2024cfk,
    author = "Navas, S. and others",
    collaboration = "Particle Data Group",
    title = "{Review of particle physics}",
    doi = "10.1103/PhysRevD.110.030001",
    journal = "Phys. Rev. D",
    volume = "110",
    number = "3",
    pages = "030001",
    year = "2024"
}

@misc{GWOSC:O4a,
  title        = {O4a Open Data Release},
  howpublished = {\url{https://gwosc.org/O4/O4a/}},
  note         = {Accessed: \today},
  year         = {2025},
  organization = {GW Open Science Center},
}

@techreport{GoetzRiles:T2400058:v1,
  author       = {Evan Goetz and Keith Riles},
  title        = {Segments used for creating standard SFTs in O4 data},
  type         = {Technical Note},
  institution  = {LIGO Laboratory},
  number       = {LIGO-T2400058-v1},
  year         = {2025},
  note         = {Version v1; other version: LIGO-T2400058-v2},
  url          = {https://dcc.ligo.org/T2400058-v1/public},
}

@software{DCHgate2025,
  title        = {{DCHgate}},
  author       = {{LIGO Scientific Collaboration, Virgo and KAGRA}},
  year         = {2025},
  url          = {https://git.ligo.org/detchar/tools/dchgate},
  organization = {LIGO Scientific Collaboration}
}

@techreport{DavisEtAl2024_selfgating_O4a,
  title        = {Self-gating of O4a \(h(t)\) for use in continuous-wave searches},
  author       = {Derek Davis and Ansel Neunzert and Evan Goetz and Keith Riles and Karl Wette and Max Lalleman},
  institution  = {LIGO Document Control Center (DCC)},
  type         = {Technical Report},
  number       = {T2400003},
  year         = {2024},
  month        = jan,
  url          = {https://dcc.ligo.org/T2400003/},
  note         = {Document created 07 Jan 2024; contents revised 23 Jan 2024},
}

@article{Marzola:2017lbt,
    author = "Marzola, Luca and Raidal, Martti and Urban, Federico R.",
    title = "{Oscillating Spin-2 Dark Matter}",
    eprint = "1708.04253",
    archivePrefix = "arXiv",
    primaryClass = "hep-ph",
    doi = "10.1103/PhysRevD.97.024010",
    journal = "Phys. Rev. D",
    volume = "97",
    number = "2",
    pages = "024010",
    year = "2018"
}

@article{Aoki:2016zgp,
    author = "Aoki, Katsuki and Mukohyama, Shinji",
    title = "{Massive gravitons as dark matter and gravitational waves}",
    eprint = "1604.06704",
    archivePrefix = "arXiv",
    primaryClass = "hep-th",
    doi = "10.1103/PhysRevD.94.024001",
    journal = "Phys. Rev. D",
    volume = "94",
    number = "2",
    pages = "024001",
    year = "2016"
}

@article{Feldman:1997qc,
    author = "Feldman, Gary J. and Cousins, Robert D.",
    title = "{A Unified approach to the classical statistical analysis of small signals}",
    eprint = "physics/9711021",
    archivePrefix = "arXiv",
    reportNumber = "HUTP-97-A096",
    doi = "10.1103/PhysRevD.57.3873",
    journal = "Phys. Rev. D",
    volume = "57",
    pages = "3873--3889",
    year = "1998"
}

@article{Schlamminger:2007ht,
    author = "Schlamminger, Stephan and Choi, K. -Y. and Wagner, T. A. and Gundlach, J. H. and Adelberger, E. G.",
    title = "{Test of the equivalence principle using a rotating torsion balance}",
    eprint = "0712.0607",
    archivePrefix = "arXiv",
    primaryClass = "gr-qc",
    doi = "10.1103/PhysRevLett.100.041101",
    journal = "Phys. Rev. Lett.",
    volume = "100",
    pages = "041101",
    year = "2008"
}

@article{Berge:2017ovy,
    author = "Berg\'e, Joel and Brax, Philippe and M\'etris, Gilles and Pernot-Borr\`as, Martin and Touboul, Pierre and Uzan, Jean-Philippe",
    title = "{MICROSCOPE Mission: First Constraints on the Violation of the Weak Equivalence Principle by a Light Scalar Dilaton}",
    eprint = "1712.00483",
    archivePrefix = "arXiv",
    primaryClass = "gr-qc",
    doi = "10.1103/PhysRevLett.120.141101",
    journal = "Phys. Rev. Lett.",
    volume = "120",
    number = "14",
    pages = "141101",
    year = "2018"
}

@article{Bertone:2016nfn,
	title        = {{History of dark matter}},
	author       = {Bertone, Gianfranco and Hooper, Dan},
	year         = 2018,
	journal      = {Rev. Mod. Phys.},
	volume       = 90,
	number       = 4,
	pages        = {045002},
	doi          = {10.1103/RevModPhys.90.045002},
	eprint       = {1605.04909},
	archiveprefix = {arXiv},
	primaryclass = {astro-ph.CO},
	reportnumber = {FERMILAB-PUB-16-157-A}
}

@article{Read:2014qva,
    author = "Read, J. I.",
    title = "{The Local Dark Matter Density}",
    eprint = "1404.1938",
    archivePrefix = "arXiv",
    primaryClass = "astro-ph.GA",
    reportNumber = "JPHYSG-100038.R1",
    doi = "10.1088/0954-3899/41/6/063101",
    journal = "J. Phys. G",
    volume = "41",
    pages = "063101",
    year = "2014"
}

@mastersthesis{rellathesis,
author = {Rella, Vincenzo},
year = {2021},
month = {09},
pages = {},
title = {Searching for ultra-light dark matter with gravitational-wave interferometers},
doi = {10.13140/RG.2.2.18029.29924}
}

@article{Riles:2022wwz,
	title        = {{Searches for continuous-wave gravitational radiation}},
	author       = {Riles, Keith},
	year         = 2023,
	journal      = {Living Rev. Rel.},
	volume       = 26,
	number       = 1,
	pages        = 3,
	doi          = {10.1007/s41114-023-00044-3},
	eprint       = {2206.06447},
	archiveprefix = {arXiv},
	primaryclass = {astro-ph.HE}
}

@article{Morisaki:2020gui,
	title        = {{Improved sensitivity of interferometric gravitational wave detectors to ultralight vector dark matter from the finite light-traveling time}},
	author       = {Morisaki, Soichiro and Fujita, Tomohiro and Michimura, Yuta and Nakatsuka, Hiromasa and Obata, Ippei},
	year         = 2021,
	journal      = {Phys. Rev. D},
	volume       = 103,
	number       = 5,
	pages        = {L051702},
	doi          = {10.1103/PhysRevD.103.L051702},
	eprint       = {2011.03589},
	archiveprefix = {arXiv},
	primaryclass = {hep-ph}
}

@article{Ferreira:2020fam,
    author = "Ferreira, Elisa G. M.",
    title = "{Ultra-light dark matter}",
    eprint = "2005.03254",
    archivePrefix = "arXiv",
    primaryClass = "astro-ph.CO",
    doi = "10.1007/s00159-021-00135-6",
    journal = "Astron. Astrophys. Rev.",
    volume = "29",
    number = "1",
    pages = "7",
    year = "2021"
}

@article{Nguyen:2024fpq,
    author = "Nguyen, Quynh Lan and Miller, Andrew L.",
    title = "{Dark Matter and its Effect on Gravitational Wave Signal}",
    doi = "10.22323/1.449.0132",
    journal = "PoS",
    volume = "EPS-HEP2023",
    pages = "132",
    year = "2024"
}

@article{gw170817FIRST,
	title        = {{GW170817}: Observation of Gravitational Waves from a Binary Neutron Star Inspiral},
	author       = {Abbott, B. P. and others},
	year         = 2017,
	month        = {Oct},
	journal      = {Physical Review Letters},
	publisher    = {American Physical Society},
	volume       = 119,
	pages        = 161101,
	doi          = {10.1103/PhysRevLett.119.161101},
	url          = {https://link.aps.org/doi/10.1103/PhysRevLett.119.161101},
	collaboration = {LIGO Scientific Collaboration and Virgo Collaboration},
	issue        = 16,
	numpages     = 18
}

@article{Aoki:2017cnz,
	title        = {{Condensate of Massive Graviton and Dark Matter}},
	author       = {Aoki, Katsuki and Maeda, Kei-ichi},
	year         = 2018,
	journal      = {Phys. Rev. D},
	volume       = 97,
	number       = 4,
	pages        = {044002},
	doi          = {10.1103/PhysRevD.97.044002},
	eprint       = {1707.05003},
	archiveprefix = {arXiv},
	primaryclass = {hep-th},
	reportnumber = {WU-AP-1702-17}
}

@article{Armaleo:2020yml,
	title        = {{Pulsar timing array constraints on spin-2 ULDM}},
	author       = {Armaleo, Juan Manuel and L\'opez Nacir, Diana and Urban, Federico R.},
	year         = 2020,
	journal      = {JCAP},
	volume       = {09},
	pages        = {031},
	doi          = {10.1088/1475-7516/2020/09/031},
	eprint       = {2005.03731},
	archiveprefix = {arXiv},
	primaryclass = {astro-ph.CO}
}

@article{LIGOScientific:2021djp,
	title        = {{GWTC-3: Compact Binary Coalescences Observed by LIGO and Virgo During the Second Part of the Third Observing Run}},
	author       = {Abbott, R. and others},
	year         = 2021,
	month        = 11,
	collaboration = {LIGO Scientific Collaboration, Virgo, KAGRA},
	eprint       = {2111.03606},
	archiveprefix = {arXiv},
	primaryclass = {gr-qc},
	reportnumber = {LIGO-P2000318}
}

@article{Astone:2014esa,
	title        = {Method for all-sky searches of continuous gravitational wave signals using the {Frequency-Hough} transform},
	author       = {Astone, Pia and Colla, Alberto and D'Antonio, Sabrina and Frasca, Sergio and Palomba, Cristiano},
	year         = 2014,
	journal      = {Physical Review D},
	publisher    = {APS},
	volume       = 90,
	number       = 4,
	pages        = {042002}
}

@article{KAGRA:2020tym,
	title        = {{Overview of KAGRA: Detector design and construction history}},
	author       = {Akutsu, T. and others},
	year         = 2021,
	journal      = {PTEP},
	volume       = 2021,
	number       = 5,
	pages        = {05A101},
	doi          = {10.1093/ptep/ptaa125},
	collaboration = {KAGRA},
	eprint       = {2005.05574},
	archiveprefix = {arXiv},
	primaryclass = {physics.ins-det}
}

@article{Aiello:2021wlp,
	title        = {{Constraints on Scalar Field Dark Matter from Colocated Michelson Interferometers}},
	author       = {Aiello, Lorenzo and Richardson, Jonathan W. and Vermeulen, Sander M. and Grote, Hartmut and Hogan, Craig and Kwon, Ohkyung and Stoughton, Chris},
	year         = 2022,
	journal      = {Phys. Rev. Lett.},
	volume       = 128,
	number       = 12,
	pages        = 121101,
	doi          = {10.1103/PhysRevLett.128.121101},
	eprint       = {2108.04746},
	archiveprefix = {arXiv},
	primaryclass = {gr-qc},
	reportnumber = {FERMILAB-PUB-21-352-SCD-T}
}

@article{Aharony:2019iad,
	title        = {{Constraining Rapidly Oscillating Scalar Dark Matter Using Dynamic Decoupling}},
	author       = {Aharony, Shahaf and Akerman, Nitzan and Ozeri, Roee and Perez, Gilad and Savoray, Inbar and Shaniv, Ravid},
	year         = 2021,
	journal      = {Phys. Rev. D},
	volume       = 103,
	number       = 7,
	pages        = {075017},
	doi          = {10.1103/PhysRevD.103.075017},
	eprint       = {1902.02788},
	archiveprefix = {arXiv},
	primaryclass = {hep-ph}
}

@article{Savalle:2020vgz,
	title        = {{Searching for Dark Matter with an Optical Cavity and an Unequal-Delay Interferometer}},
	author       = {Savalle, Etienne and Hees, Aur\'elien and Frank, Florian and Cantin, Etienne and Pottie, Paul-Eric and Roberts, Benjamin M. and Cros, Lucie and Mcallister, Ben T. and Wolf, Peter},
	year         = 2021,
	journal      = {Phys. Rev. Lett.},
	volume       = 126,
	number       = 5,
	pages        = {051301},
	doi          = {10.1103/PhysRevLett.126.051301},
	eprint       = {2006.07055},
	archiveprefix = {arXiv},
	primaryclass = {gr-qc}
}

@article{Kennedy:2020bac,
	title        = {{Precision Metrology Meets Cosmology: Improved Constraints on Ultralight Dark Matter from Atom-Cavity Frequency Comparisons}},
	author       = {Kennedy, Colin J. and Oelker, Eric and Robinson, John M. and Bothwell, Tobias and Kedar, Dhruv and Milner, William R. and Marti, G. Edward and Derevianko, Andrei and Ye, Jun},
	year         = 2020,
	journal      = {Phys. Rev. Lett.},
	volume       = 125,
	number       = 20,
	pages        = 201302,
	doi          = {10.1103/PhysRevLett.125.201302},
	eprint       = {2008.08773},
	archiveprefix = {arXiv},
	primaryclass = {physics.atom-ph}
}

@article{Antypas:2019qji,
	title        = {{Scalar dark matter in the radio-frequency band: atomic-spectroscopy search results}},
	author       = {Antypas, D. and Tretiak, O. and Garcon, A. and Ozeri, R. and Perez, G. and Budker, D.},
	year         = 2019,
	journal      = {Phys. Rev. Lett.},
	volume       = 123,
	number       = 14,
	pages        = 141102,
	doi          = {10.1103/PhysRevLett.123.141102},
	eprint       = {1905.02968},
	archiveprefix = {arXiv},
	primaryclass = {physics.atom-ph}
}

@article{Antypas:2020rtg,
	title        = {{Probing fast oscillating scalar dark matter with atoms and molecules}},
	author       = {Antypas, Dionysios and Tretiak, Oleg and Zhang, Ke and Garcon, Antoine and Perez, Gilad and Kozlov, Mikhail G. and Schiller, Stephan and Budker, Dmitry},
	year         = 2021,
	journal      = {Quantum Sci. Technol.},
	volume       = 6,
	number       = 3,
	pages        = {034001},
	doi          = {10.1088/2058-9565/abe472},
	eprint       = {2012.01519},
	archiveprefix = {arXiv},
	primaryclass = {physics.atom-ph}
}

@article{Tretiak:2022ndx,
	title        = {{Improved Bounds on Ultralight Scalar Dark Matter in the Radio-Frequency Range}},
	author       = {Tretiak, Oleg and Zhang, Xue and Figueroa, Nataniel L. and Antypas, Dionysios and Brogna, Andrea and Banerjee, Abhishek and Perez, Gilad and Budker, Dmitry},
	year         = 2022,
	journal      = {Phys. Rev. Lett.},
	volume       = 129,
	number       = 3,
	pages        = {031301},
	doi          = {10.1103/PhysRevLett.129.031301},
	eprint       = {2201.02042},
	archiveprefix = {arXiv},
	primaryclass = {hep-ph}
}

@article{Oswald:2021vtc,
	title        = {{Search for Dark-Matter-Induced Oscillations of Fundamental Constants Using Molecular Spectroscopy}},
	author       = {Oswald, R. and others},
	year         = 2022,
	journal      = {Phys. Rev. Lett.},
	volume       = 129,
	number       = 3,
	pages        = {031302},
	doi          = {10.1103/PhysRevLett.129.031302},
	eprint       = {2111.06883},
	archiveprefix = {arXiv},
	primaryclass = {hep-ph}
}

@article{Campbell:2020fvq,
	title        = {{Searching for Scalar Dark Matter via Coupling to Fundamental Constants with Photonic, Atomic and Mechanical Oscillators}},
	author       = {Campbell, William M. and McAllister, Ben T. and Goryachev, Maxim and Ivanov, Eugene N. and Tobar, Michael E.},
	year         = 2021,
	journal      = {Phys. Rev. Lett.},
	volume       = 126,
	number       = 7,
	pages        = {071301},
	doi          = {10.1103/PhysRevLett.126.071301},
	eprint       = {2010.08107},
	archiveprefix = {arXiv},
	primaryclass = {hep-ex}
}

@article{BACON:2020ubh,
	title        = {{Frequency ratio measurements at 18-digit accuracy using an optical clock network}},
	author       = {Beloy, Kyle and others},
	year         = 2021,
	journal      = {Nature},
	volume       = 591,
	number       = 7851,
	pages        = {564--569},
	doi          = {10.1038/s41586-021-03253-4},
	collaboration = {BACON},
	eprint       = {2005.14694},
	archiveprefix = {arXiv},
	primaryclass = {physics.atom-ph}
}

@article{Zhang:2022ewz,
	title        = {{Search for Ultralight Dark Matter with Spectroscopy of Radio-Frequency Atomic Transitions}},
	author       = {Zhang, Xue and Banerjee, Abhishek and Leyser, Mahapan and Perez, Gilad and Schiller, Stephan and Budker, Dmitry and Antypas, Dionysios},
	year         = 2023,
	journal      = {Phys. Rev. Lett.},
	volume       = 130,
	number       = 25,
	pages        = 251002,
	doi          = {10.1103/PhysRevLett.130.251002},
	eprint       = {2212.04413},
	archiveprefix = {arXiv},
	primaryclass = {physics.atom-ph}
}

@article{Fukusumi:2023kqd,
	title        = {{Upper limit on scalar field dark matter from LIGO-Virgo third observation run}},
	author       = {Fukusumi, Koki and Morisaki, Soichiro and Suyama, Teruaki},
	year         = 2023,
	journal      = {Phys. Rev. D},
	volume       = 108,
	number       = 9,
	pages        = {095054},
	doi          = {10.1103/PhysRevD.108.095054},
	eprint       = {2303.13088},
	archiveprefix = {arXiv},
	primaryclass = {hep-ph}
}

@article{Sherrill:2023zah,
	title        = {{Analysis of atomic-clock data to constrain variations of fundamental constants}},
	author       = {Sherrill, Nathaniel and others},
	year         = 2023,
	journal      = {New J. Phys.},
	volume       = 25,
	number       = 9,
	pages        = {093012},
	doi          = {10.1088/1367-2630/aceff6},
	eprint       = {2302.04565},
	archiveprefix = {arXiv},
	primaryclass = {physics.atom-ph}
}

@article{Manita:2023mnc,
	title        = {{Exploring the spin of ultralight dark matter with gravitational wave detectors}},
	author       = {Manita, Yusuke and Takeda, Hiroki and Aoki, Katsuki and Fujita, Tomohiro and Mukohyama, Shinji},
	year         = 2024,
	month        = 10,
	journal      = {Phys. Rev. D},
	volume       = 109,
	number       = 9,
	pages        = {095012},
	doi          = {10.1103/PhysRevD.109.095012},
	eprint       = {2310.10646},
	archiveprefix = {arXiv},
	primaryclass = {hep-ph},
	reportnumber = {KUNS-2979, YITP-23-122, IPMU23-0034}
}

@article{Piccinni:2018akm,
	title        = {{A new data analysis framework for the search of continuous gravitational wave signals}},
	author       = {Piccinni, O. J. and Astone, P. and D'Antonio, S. and Frasca, S. and Intini, G. and Leaci, P. and Mastrogiovanni, S. and Miller, A. and Palomba, C. and Singhal, A.},
	year         = 2019,
	journal      = {Class. Quant. Grav.},
	volume       = 36,
	number       = 1,
	pages        = {015008},
	doi          = {10.1088/1361-6382/aaefb5},
	eprint       = {1811.04730},
	archiveprefix = {arXiv},
	primaryclass = {gr-qc}
}

@article{LIGO:2024kkz,
    author = "Soni, S. and others",
    collaboration = "LIGO",
    title = "{LIGO Detector Characterization in the first half of the fourth Observing run}",
    eprint = "2409.02831",
    archivePrefix = "arXiv",
    primaryClass = "astro-ph.IM",
    doi = "10.1088/1361-6382/adc4b6",
    journal = "Class. Quant. Grav.",
    volume = "42",
    number = "8",
    pages = "085016",
    year = "2025"
}

@article{Astone:2005fj,
	title        = {{The short FFT database and the peak map for the hierarchical search of periodic sources}},
	author       = {Astone, P. and Frasca, S. and Palomba, C.},
	year         = 2005,
	journal      = {Class. Quant. Grav.},
	volume       = 22,
	pages        = {S1197--S1210},
	doi          = {10.1088/0264-9381/22/18/S34},
	editor       = {Mours, B. and Marion, F.}
}

@article{Sereno:2006mw,
    author = "Sereno, Mauro and Jetzer, Ph.",
    title = "{Dark matter vs. modifications of the gravitational inverse-square law. Results from planetary motion in the solar system}",
    eprint = "astro-ph/0606197",
    archivePrefix = "arXiv",
    doi = "10.1111/j.1365-2966.2006.10670.x",
    journal = "Mon. Not. Roy. Astron. Soc.",
    volume = "371",
    pages = "626--632",
    year = "2006"
}

@article{Murata:2014nra,
    author = "Murata, Jiro and Tanaka, Saki",
    title = "{A review of short-range gravity experiments in the LHC era}",
    eprint = "1408.3588",
    archivePrefix = "arXiv",
    primaryClass = "hep-ex",
    doi = "10.1088/0264-9381/32/3/033001",
    journal = "Class. Quant. Grav.",
    volume = "32",
    number = "3",
    pages = "033001",
    year = "2015"
}

@book{maggiore2008gravitational,
	title        = {Gravitational Waves: Volume 1: Theory and Experiments},
	author       = {Maggiore, Michele},
	year         = 2008,
	publisher    = {Oxford University Press},
	volume       = 1
}

@article{Bertone:2019irm,
	title        = {{Gravitational wave probes of dark matter: challenges and opportunities}},
	author       = {Bertone, Gianfranco and others},
	year         = 2020,
	journal      = {SciPost Phys. Core},
	volume       = 3,
	pages        = {007},
	doi          = {10.21468/SciPostPhysCore.3.2.007},
	eprint       = {1907.10610},
	archiveprefix = {arXiv},
	primaryclass = {astro-ph.CO}
}

@article{Guo:2019ker,
	title        = {{Searching for Dark Photon Dark Matter in LIGO O1 Data}},
	author       = {Guo, Huai-Ke and Riles, Keith and Yang, Feng-Wei and Zhao, Yue},
	year         = 2019,
	journal      = {Commun. Phys.},
	volume       = 2,
	pages        = 155,
	doi          = {10.1038/s42005-019-0255-0},
	eprint       = {1905.04316},
	archiveprefix = {arXiv},
	primaryclass = {hep-ph}
}

@article{Co:2018lka,
	title        = {{Dark Photon Dark Matter Produced by Axion Oscillations}},
	author       = {Co, Raymond T. and Pierce, Aaron and Zhang, Zhengkang and Zhao, Yue},
	year         = 2019,
	journal      = {Phys. Rev. D},
	volume       = 99,
	number       = 7,
	pages        = {075002},
	doi          = {10.1103/PhysRevD.99.075002},
	eprint       = {1810.07196},
	archiveprefix = {arXiv},
	primaryclass = {hep-ph},
	reportnumber = {LCTP-18-21}
}

@article{Dror:2018pdh,
	title        = {{Parametric Resonance Production of Ultralight Vector Dark Matter}},
	author       = {Dror, Jeff A. and Harigaya, Keisuke and Narayan, Vijay},
	year         = 2019,
	journal      = {Phys. Rev. D},
	volume       = 99,
	number       = 3,
	pages        = {035036},
	doi          = {10.1103/PhysRevD.99.035036},
	eprint       = {1810.07195},
	archiveprefix = {arXiv},
	primaryclass = {hep-ph}
}

@article{Bastero-Gil:2018uel,
	title        = {{Vector dark matter production at the end of inflation}},
	author       = {Bastero-Gil, Mar and Santiago, Jose and Ubaldi, Lorenzo and Vega-Morales, Roberto},
	year         = 2019,
	journal      = {JCAP},
	volume       = {04},
	pages        = {015},
	doi          = {10.1088/1475-7516/2019/04/015},
	eprint       = {1810.07208},
	archiveprefix = {arXiv},
	primaryclass = {hep-ph},
	reportnumber = {UG-FT 328-18, CAFPE 198-18, SISSA 44/2018/FISI}
}

@article{Holdom:1985ag,
	title        = {{Two U(1)'s and Epsilon Charge Shifts}},
	author       = {Holdom, Bob},
	year         = 1986,
	journal      = {Phys. Lett. B},
	volume       = 166,
	pages        = {196--198},
	doi          = {10.1016/0370-2693(86)91377-8},
	reportnumber = {UTPT-85-30}
}

@article{Du:2018uak,
	title        = {{A Search for Invisible Axion Dark Matter with the Axion Dark Matter Experiment}},
	author       = {Du, N. and others},
	year         = 2018,
	journal      = {Phys. Rev. Lett.},
	volume       = 120,
	number       = 15,
	pages        = 151301,
	doi          = {10.1103/PhysRevLett.120.151301},
	collaboration = {ADMX},
	eprint       = {1804.05750},
	archiveprefix = {arXiv},
	primaryclass = {hep-ex},
	reportnumber = {FERMILAB-PUB-18-101-AD-AE}
}

@article{Hartman:2020bgj,
	title        = {{Design of the ALPS II Optical System}},
	author       = {Ortiz, M. Diaz and others},
	year         = 2020,
	month        = 9,
	eprint       = {2009.14294},
	archiveprefix = {arXiv},
	primaryclass = {physics.optics},
	reportnumber = {DESY 20-158, DESY-20-158}
}

@article{Su:1994gu,
	title        = {{New tests of the universality of free fall}},
	author       = {Su, Y. and Heckel, Blayne R. and Adelberger, E. G. and Gundlach, J. H. and Harris, M. and Smith, G. L. and Swanson, H. E.},
	year         = 1994,
	journal      = {Phys. Rev. D},
	volume       = 50,
	pages        = {3614--3636},
	doi          = {10.1103/PhysRevD.50.3614}
}

@article{Touboul:2012ui,
	title        = {{The MICROSCOPE experiment, ready for the in-orbit test of the equivalence principle}},
	author       = {Touboul, P. and Metris, G. and Lebat, V. and Robert, A.},
	year         = 2012,
	journal      = {Class. Quant. Grav.},
	volume       = 29,
	pages        = 184010,
	doi          = {10.1088/0264-9381/29/18/184010}
}

@article{Pierce:2018xmy,
	title        = {{Searching for Dark Photon Dark Matter with Gravitational Wave Detectors}},
	author       = {Pierce, Aaron and Riles, Keith and Zhao, Yue},
	year         = 2018,
	journal      = {Phys. Rev. Lett.},
	volume       = 121,
	number       = 6,
	pages        = {061102},
	doi          = {10.1103/PhysRevLett.121.061102},
	eprint       = {1801.10161},
	archiveprefix = {arXiv},
	primaryclass = {hep-ph},
	reportnumber = {LCTP-18-04}
}

@article{Agrawal:2018vin,
    author = "Agrawal, Prateek and Kitajima, Naoya and Reece, Matthew and Sekiguchi, Toyokazu and Takahashi, Fuminobu",
    title = "{Relic Abundance of Dark Photon Dark Matter}",
    eprint = "1810.07188",
    archivePrefix = "arXiv",
    primaryClass = "hep-ph",
    reportNumber = "TU-1074,IPMU18-015,RESCEU-13/18,MIT-CTP/5066, TU-1074, IPMU18-015, RESCEU-13/18, MIT-CTP/5066",
    doi = "10.1016/j.physletb.2019.135136",
    journal = "Phys. Lett. B",
    volume = "801",
    pages = "135136",
    year = "2020"
}

@article{long2019dark,
	title        = {Dark photon dark matter from a network of cosmic strings},
	author       = {Long, Andrew J and Wang, Lian-Tao},
	year         = 2019,
	journal      = {Physical Review D},
	publisher    = {APS},
	volume       = 99,
	number       = 6,
	pages        = {063529}
}

@article{nelson2011dark,
	title        = {Dark light, dark matter, and the misalignment mechanism},
	author       = {Nelson, Ann E and Scholtz, Jakub},
	year         = 2011,
	journal      = {Physical Review D},
	publisher    = {APS},
	volume       = 84,
	number       = 10,
	pages        = 103501
}

@article{agrawal2020relic,
	title        = {Relic abundance of dark photon dark matter},
	author       = {Agrawal, Prateek and Kitajima, Naoya and Reece, Matthew and Sekiguchi, Toyokazu and Takahashi, Fuminobu},
	year         = 2020,
	journal      = {Physics Letters B},
	publisher    = {Elsevier},
	volume       = 801,
	pages        = 135136
}

@article{graham2016vector,
	title        = {Vector dark matter from inflationary fluctuations},
	author       = {Graham, Peter W and Mardon, Jeremy and Rajendran, Surjeet},
	year         = 2016,
	journal      = {Physical Review D},
	publisher    = {APS},
	volume       = 93,
	number       = 10,
	pages        = 103520
}

@article{arias2012wispy,
	title        = {WISPy cold dark matter},
	author       = {Arias, Paola and Cadamuro, Davide and Goodsell, Mark and Jaeckel, Joerg and Redondo, Javier and Ringwald, Andreas},
	year         = 2012,
	journal      = {Journal of Cosmology and Astroparticle Physics},
	publisher    = {IOP Publishing},
	volume       = 2012,
	number       = {06},
	pages        = {013}
}

@article{Arvanitaki:2014wva,
	title        = {{Discovering the QCD Axion with Black Holes and Gravitational Waves}},
	author       = {Arvanitaki, Asimina and Baryakhtar, Masha and Huang, Xinlu},
	year         = 2015,
	journal      = {Phys. Rev. D},
	volume       = 91,
	number       = 8,
	pages        = {084011},
	doi          = {10.1103/PhysRevD.91.084011},
	eprint       = {1411.2263},
	archiveprefix = {arXiv},
	primaryclass = {hep-ph}
}

@article{Miller:2020vsl,
	title        = {{Probing new light gauge bosons with gravitational-wave interferometers using an adapted semicoherent method}},
	author       = {Miller, Andrew L. and others},
	year         = 2021,
	journal      = {Phys. Rev. D},
	volume       = 103,
	number       = 10,
	pages        = 103002,
	doi          = {10.1103/PhysRevD.103.103002},
	eprint       = {2010.01925},
	archiveprefix = {arXiv},
	primaryclass = {astro-ph.IM}
}

@article{Grote:2019uvn,
    author = "Grote, H. and Stadnik, Y. V.",
    title = "{Novel signatures of dark matter in laser-interferometric gravitational-wave detectors}",
    eprint = "1906.06193",
    archivePrefix = "arXiv",
    primaryClass = "astro-ph.IM",
    doi = "10.1103/PhysRevResearch.1.033187",
    journal = "Phys. Rev. Res.",
    volume = "1",
    number = "3",
    pages = "033187",
    year = "2019"
}

@article{Capote:2024rmo,
    author = "Capote, E. and others",
    title = "{Advanced LIGO detector performance in the fourth observing run}",
    eprint = "2411.14607",
    archivePrefix = "arXiv",
    primaryClass = "gr-qc",
    reportNumber = "LIGO-P2400256",
    doi = "10.1103/PhysRevD.111.062002",
    journal = "Phys. Rev. D",
    volume = "111",
    number = "6",
    pages = "062002",
    year = "2025"
}

@article{LIGOO4Detector:2023wmz,
    author = "Ganapathy, D. and others",
    collaboration = "LIGO O4 Detector",
    title = "{Broadband Quantum Enhancement of the LIGO Detectors with Frequency-Dependent Squeezing}",
    doi = "10.1103/PhysRevX.13.041021",
    journal = "Phys. Rev. X",
    volume = "13",
    number = "4",
    pages = "041021",
    year = "2023"
}

@article{membersoftheLIGOScientific:2024elc,
    author = "Jia, Wenxuan and others",
    collaboration = "members of the LIGO Scientific{\textdagger}",
    title = "{Squeezing the quantum noise of a gravitational-wave detector below the standard quantum limit}",
    eprint = "2404.14569",
    archivePrefix = "arXiv",
    primaryClass = "gr-qc",
    reportNumber = "LIGO-P2400059",
    doi = "10.1126/science.ado8069",
    journal = "Science",
    volume = "385",
    number = "6715",
    pages = "1318",
    year = "2024"
}

@article{Miller:2025yyx,
    author = "Miller, Andrew L.",
    title = "{Gravitational wave probes of particle dark matter: A review}",
    eprint = "2503.02607",
    archivePrefix = "arXiv",
    primaryClass = "astro-ph.HE",
    doi = "10.1142/S0218271825300058",
    journal = "Int. J. Mod. Phys. D",
    volume = "35",
    number = "01",
    pages = "2530005",
    year = "2026"
}

@article{PhysRevD.100.123512,
	title        = {Detectability of ultralight scalar field dark matter with gravitational-wave detectors},
	author       = {Morisaki, Soichiro and Suyama, Teruaki},
	year         = 2019,
	month        = {Dec},
	journal      = {Phys. Rev. D},
	publisher    = {American Physical Society},
	volume       = 100,
	pages        = 123512,
	doi          = {10.1103/PhysRevD.100.123512},
	url          = {https://link.aps.org/doi/10.1103/PhysRevD.100.123512},
	issue        = 12,
	numpages     = 14
}

@article{Stadnik2015a,
	title        = {{Searching for Dark Matter and Variation of Fundamental Constants with Laser and Maser Interferometry}},
	author       = {Stadnik, Yevgeny and Flambaum, Victor},
	year         = 2015,
	journal      = {Physical Review Letters},
	volume       = 114,
	pages        = 161301
}

@article{LIGOScientific:2025bkz,
    author = "Abac, A. G. and others",
    collaboration = "LIGO Scientific, VIRGO, KAGRA",
    title = "{Directional Search for Persistent Gravitational Waves: Results from the First Part of LIGO-Virgo-KAGRA's Fourth Observing Run}",
    eprint = "2510.17487",
    archivePrefix = "arXiv",
    primaryClass = "gr-qc",
    reportNumber = "LIGO-P250038",
    month = "10",
    year = "2025"
}

@article{LIGOScientific:2025bgj,
    author = "Abac, A. G. and others",
    collaboration = "LIGO Scientific, VIRGO, KAGRA",
    title = "{Upper Limits on the Isotropic Gravitational-Wave Background from the first part of LIGO, Virgo, and KAGRA's fourth Observing Run}",
    eprint = "2508.20721",
    archivePrefix = "arXiv",
    primaryClass = "gr-qc",
    reportNumber = "LIGO-P2500349",
    month = "8",
    year = "2025"
}

@article{Ng:2019gch,
    author = "Ng, Kenny C. Y. and Roach, Brandon M. and Perez, Kerstin and Beacom, John F. and Horiuchi, Shunsaku and Krivonos, Roman and Wik, Daniel R.",
    title = "{New Constraints on Sterile Neutrino Dark Matter from $NuSTAR$ M31 Observations}",
    eprint = "1901.01262",
    archivePrefix = "arXiv",
    primaryClass = "astro-ph.HE",
    doi = "10.1103/PhysRevD.99.083005",
    journal = "Phys. Rev. D",
    volume = "99",
    pages = "083005",
    year = "2019"
}

@article{PerezdelosHeros:2020qyt,
    author = "P{\'e}rez de los Heros, Carlos",
    title = "{Status, Challenges and Directions in Indirect Dark Matter Searches}",
    eprint = "2008.11561",
    archivePrefix = "arXiv",
    primaryClass = "astro-ph.HE",
    doi = "10.3390/sym12101648",
    journal = "Symmetry",
    volume = "12",
    number = "10",
    pages = "1648",
    year = "2020"
}

@article{Stadnik2015b,
	title        = {{Can Dark Matter Induce Cosmological Evolution of the Fundamental Constants of Nature?}},
	author       = {Stadnik, Yevgeny and Flambaum, Victor},
	year         = 2015,
	journal      = {Physical Review Letters},
	volume       = 115,
	pages        = 201301
}

@article{Preskill:1982cy,
	title        = {{Cosmology of the Invisible Axion}},
	author       = {Preskill, John and Wise, Mark B. and Wilczek, Frank},
	year         = 1983,
	journal      = {Phys. Lett. B},
	volume       = 120,
	pages        = {127--132},
	doi          = {10.1016/0370-2693(83)90637-8},
	editor       = {Srednicki, M.A.},
	reportnumber = {HUTP-82-A048, NSF-ITP-82-103}
}

@article{Abbott:1982af,
	title        = {{A Cosmological Bound on the Invisible Axion}},
	author       = {Abbott, L.F. and Sikivie, P.},
	year         = 1983,
	journal      = {Phys. Lett. B},
	volume       = 120,
	pages        = {133--136},
	doi          = {10.1016/0370-2693(83)90638-X},
	editor       = {Srednicki, M.A.},
	reportnumber = {PRINT-82-0695 (BRANDEIS)}
}

@article{Hassan:2011zd,
	title        = {{Bimetric Gravity from Ghost-free Massive Gravity}},
	author       = {Hassan, S. F. and Rosen, Rachel A.},
	year         = 2012,
	journal      = {JHEP},
	volume       = {02},
	pages        = 126,
	doi          = {10.1007/JHEP02(2012)126},
	eprint       = {1109.3515},
	archiveprefix = {arXiv},
	primaryclass = {hep-th}
}

@article{Cho:2007cy,
	title        = {{Dilatonic dark matter and its experimental detection}},
	author       = {Cho, Y.M. and Kim, J.H.},
	year         = 2009,
	journal      = {Phys. Rev. D},
	volume       = 79,
	pages        = {023504},
	doi          = {10.1103/PhysRevD.79.023504},
	eprint       = {0711.2858},
	archiveprefix = {arXiv},
	primaryclass = {gr-qc}
}

@article{Arvanitaki:2014faa,
	title        = {{Searching for dilaton dark matter with atomic clocks}},
	author       = {Arvanitaki, Asimina and Huang, Junwu and Van Tilburg, Ken},
	year         = 2015,
	journal      = {Phys. Rev. D},
	volume       = 91,
	number       = 1,
	pages        = {015015},
	doi          = {10.1103/PhysRevD.91.015015},
	eprint       = {1405.2925},
	archiveprefix = {arXiv},
	primaryclass = {hep-ph}
}

@article{Dine:1982ah,
	title        = {{The Not So Harmless Axion}},
	author       = {Dine, Michael and Fischler, Willy},
	year         = 1983,
	journal      = {Phys. Lett. B},
	volume       = 120,
	pages        = {137--141},
	doi          = {10.1016/0370-2693(83)90639-1},
	editor       = {Srednicki, M.A.},
	reportnumber = {UPR-0201T}
}

@article{Stadnik2016,
	title        = {{Enhanced effects of variation of the fundamental constants in laser interferometers and application to dark-matter detection}},
	author       = {Stadnik, Yevgeny and Flambaum, Victor},
	year         = 2016,
	journal      = {Physical Review A},
	volume       = 93,
	pages        = {063630}
}

@book{wiener1964extrapolation,
	title        = {Extrapolation, interpolation, and smoothing of stationary time series: with engineering applications},
	author       = {Wiener, Norbert and others},
	year         = 1964,
	publisher    = {MIT press Cambridge, MA},
	volume       = 8
}

@article{Karki:2016pht,
    author = "Karki, S. and others",
    title = "{The Advanced LIGO Photon Calibrators}",
    eprint = "1608.05055",
    archivePrefix = "arXiv",
    primaryClass = "astro-ph.IM",
    doi = "10.1063/1.4967303",
    journal = "Rev. Sci. Instrum.",
    volume = "87",
    number = "11",
    pages = "114503",
    year = "2016"
}

@article{Viets:2017yvy,
    author = "Viets, Aaron and others",
    title = "{Reconstructing the calibrated strain signal in the Advanced LIGO detectors}",
    eprint = "1710.09973",
    archivePrefix = "arXiv",
    primaryClass = "astro-ph.IM",
    doi = "10.1088/1361-6382/aab658",
    journal = "Class. Quant. Grav.",
    volume = "35",
    number = "9",
    pages = "095015",
    year = "2018"
}

@article{Carr:2019kxo,
	title        = {{Cosmic conundra explained by thermal history and primordial black holes}},
	author       = {Carr, Bernard and Clesse, Sebastien and Garc\'\i{}a-Bellido, Juan and K\"uhnel, Florian},
	year         = 2021,
	journal      = {Phys. Dark Univ.},
	volume       = 31,
	pages        = 100755,
	doi          = {10.1016/j.dark.2020.100755},
	eprint       = {1906.08217},
	archiveprefix = {arXiv},
	primaryclass = {astro-ph.CO}
}

@article{2015CQGra..32g4001L,
	title        = {{Advanced LIGO}},
	author       = {{Aasi}, J. and {Abbott}, B.~P. and {Abbott}, R. and {Abbott}, T. and {Abernathy}, M.~R. and {Ackley}, K. and {Adams}, C. and {Adams}, T. and {Addesso}, P. and et al.},
	year         = 2015,
	month        = apr,
	journal      = {CQGra},
	volume       = 32,
	number       = 7,
	pages        = {074001},
	doi          = {10.1088/0264-9381/32/7/074001},
	archiveprefix = {arXiv},
	eprint       = {1411.4547},
	primaryclass = {gr-qc},
	eid          = {074001},
	adsurl       = {http://adsabs.harvard.edu/abs/2015CQGra..32g4001L}
}

@article{2015CQGra..32b4001A,
	title        = {{Advanced Virgo: a second-generation interferometric gravitational wave detector}},
	author       = {{Acernese}, F. and {Agathos}, M. and {Agatsuma}, K. and {Aisa}, D. and {Allemandou}, N. and {Allocca}, A. and {Amarni}, J. and {Astone}, P. and {Balestri}, G. and {Ballardin}, G. and et al.},
	year         = 2015,
	month        = jan,
	journal      = {CQGra},
	volume       = 32,
	number       = 2,
	pages        = {024001},
	doi          = {10.1088/0264-9381/32/2/024001},
	archiveprefix = {arXiv},
	eprint       = {1408.3978},
	primaryclass = {gr-qc},
	eid          = {024001},
	adsurl       = {http://adsabs.harvard.edu/abs/2015CQGra..32b4001A}
}

@article{Manita:2022tkl,
    author = "Manita, Yusuke and Aoki, Katsuki and Fujita, Tomohiro and Mukohyama, Shinji",
    title = "{Spin-2 dark matter from an anisotropic universe in bigravity}",
    eprint = "2211.15873",
    archivePrefix = "arXiv",
    primaryClass = "gr-qc",
    reportNumber = "KUNS-2946, YITP-22-150, IPMU22-0063",
    doi = "10.1103/PhysRevD.107.104007",
    journal = "Phys. Rev. D",
    volume = "107",
    number = "10",
    pages = "104007",
    year = "2023"
}

@article{Jain:2021pnk,
    author = "Jain, Mudit and Amin, Mustafa A.",
    title = "{Polarized solitons in higher-spin wave dark matter}",
    eprint = "2109.04892",
    archivePrefix = "arXiv",
    primaryClass = "hep-th",
    doi = "10.1103/PhysRevD.105.056019",
    journal = "Phys. Rev. D",
    volume = "105",
    number = "5",
    pages = "056019",
    year = "2022"
}

@article{LIGOScientific:2021odm,
	title        = {{Constraints on dark photon dark matter using data from LIGO's and Virgo's third observing run}},
	author       = {Abbott, R. and others},
	year         = 2022,
	journal      = {Phys. Rev. D},
	volume       = 105,
	number       = 6,
	pages        = {063030},
	doi          = {10.1103/PhysRevD.105.063030},
	collaboration = {LIGO Scientific Collaboration, Virgo, KAGRA},
	eprint       = {2105.13085},
	archiveprefix = {arXiv},
	primaryclass = {astro-ph.CO},
	reportnumber = {LIGO-P2100098}
}

@article{Armaleo:2020efr,
	title        = {{Searching for spin-2 ULDM with gravitational waves interferometers}},
	author       = {Armaleo, Juan Manuel and Nacir, Diana L\'opez and Urban, Federico R.},
	year         = 2021,
	journal      = {JCAP},
	volume       = {04},
	pages        = {053},
	doi          = {10.1088/1475-7516/2021/04/053},
	eprint       = {2012.13997},
	archiveprefix = {arXiv},
	primaryclass = {astro-ph.CO}
}

@article{Vermeulen:2021epa,
	title        = {Direct limits for scalar field dark matter from a gravitational-wave detector},
	author       = {Vermeulen, Sander M and others},
	year         = 2021,
	journal      = {Nature},
	publisher    = {Nature Publishing Group},
	volume       = 600,
	number       = 7889,
	pages        = {424--428}
}

@article{Green:2020jor,
	title        = {{Primordial Black Holes as a dark matter candidate}},
	author       = {Green, Anne M. and Kavanagh, Bradley J.},
	year         = 2021,
	journal      = {J. Phys. G},
	volume       = 48,
	number       = 4,
	pages        = {043001},
	doi          = {10.1088/1361-6471/abc534},
	eprint       = {2007.10722},
	archiveprefix = {arXiv},
	primaryclass = {astro-ph.CO}
}

@article{Miller:2022wxu,
	title        = {{Distinguishing between dark-matter interactions with gravitational-wave detectors}},
	author       = {Miller, Andrew L. and Badaracco, Francesca and Palomba, Cristiano},
	year         = 2022,
	journal      = {Phys. Rev. D},
	volume       = 105,
	number       = 10,
	pages        = 103035,
	doi          = {10.1103/PhysRevD.105.103035},
	collaboration = {LIGO Scientific Collaboration, Virgo, KAGRA},
	eprint       = {2204.03814},
	archiveprefix = {arXiv},
	primaryclass = {astro-ph.IM}
}

@misc{Guo:2025O4aDMData,
    author       = {LVK},
    title        = {{O4a Dark Matter Paper Data Product}},
    year         = {2025},
    howpublished = {LIGO Document Control Center},
    number       = {LIGO-P2500638},
    url          = {https://dcc.ligo.org/LIGO-P2500638}
}

@article{Abbott:2016blz,
	title        = {{Observation of Gravitational Waves from a Binary Black Hole Merger}},
	author       = {Abbott, B.P. and others},
	year         = 2016,
	journal      = {Phys. Rev. Lett.},
	volume       = 116,
	number       = 6,
	pages        = {061102},
	doi          = {10.1103/PhysRevLett.116.061102},
	collaboration = {LIGO Scientific Collaboration, Virgo},
	eprint       = {1602.03837},
	archiveprefix = {arXiv},
	primaryclass = {gr-qc},
	reportnumber = {LIGO-P150914}
}

@article{Carr:2016drx,
	title        = {{Primordial Black Holes as Dark Matter}},
	author       = {Carr, Bernard and Kuhnel, Florian and Sandstad, Marit},
	year         = 2016,
	journal      = {Phys. Rev. D},
	volume       = 94,
	number       = 8,
	pages        = {083504},
	doi          = {10.1103/PhysRevD.94.083504},
	eprint       = {1607.06077},
	archiveprefix = {arXiv},
	primaryclass = {astro-ph.CO},
	reportnumber = {NORDITA-2016-83}
}

@article{Gottel:2024cfj,
	title        = {{Searching for Scalar Field Dark Matter with LIGO}},
	author       = {G\"ottel, Alexandre S. and Ejlli, Aldo and Karan, Kanioar and Vermeulen, Sander M. and Aiello, Lorenzo and Raymond, Vivien and Grote, Hartmut},
	year         = 2024,
	journal      = {Phys. Rev. Lett.},
	volume       = 133,
	number       = 10,
	pages        = 101001,
	doi          = {10.1103/PhysRevLett.133.101001},
	eprint       = {2401.18076},
	archiveprefix = {arXiv},
	primaryclass = {astro-ph.CO}
}

@article{Gottel:2025aqm,
    author = {G\"ottel, Alexandre and Raymond, Vivien},
    title = "{Fast and Precise Spectral Analysis for Dark Matter Searches with LIGO}",
    eprint = "2503.03293",
    archivePrefix = "arXiv",
    primaryClass = "astro-ph.CO",
    month = "3",
    year = "2025"
}

@misc{finesse,
	title        = {Finesse},
	author       = {Brown, Daniel David and Freise, Andreas},
	year         = 2014,
	month        = may,
	doi          = {10.5281/zenodo.821363},
	url          = {http://www.gwoptics.org/finesse},
	note         = {{The software and source code is available at \url{http://www.gwoptics.org/finesse}.}}
}

@article{DeFelice:2020ecp,
    author = "De Felice, Antonio and Larrouturou, Fran{\c{c}}ois and Mukohyama, Shinji and Oliosi, Michele",
    title = "{Minimal Theory of Bigravity: construction and cosmology}",
    eprint = "2012.01073",
    archivePrefix = "arXiv",
    primaryClass = "gr-qc",
    reportNumber = "YITP-20-157, IPMU20-0126",
    doi = "10.1088/1475-7516/2021/04/015",
    journal = "JCAP",
    volume = "04",
    pages = "015",
    year = "2021"
}

@article{Khoury:2003aq,
    author = "Khoury, Justin and Weltman, Amanda",
    title = "{Chameleon fields: Awaiting surprises for tests of gravity in space}",
    eprint = "astro-ph/0309300",
    archivePrefix = "arXiv",
    doi = "10.1103/PhysRevLett.93.171104",
    journal = "Phys. Rev. Lett.",
    volume = "93",
    pages = "171104",
    year = "2004"
}

@article{Khoury:2003rn,
    author = "Khoury, Justin and Weltman, Amanda",
    title = "{Chameleon cosmology}",
    eprint = "astro-ph/0309411",
    archivePrefix = "arXiv",
    doi = "10.1103/PhysRevD.69.044026",
    journal = "Phys. Rev. D",
    volume = "69",
    pages = "044026",
    year = "2004"
}

@article{Olive:2007aj,
    author = "Olive, Keith A. and Pospelov, Maxim",
    title = "{Environmental dependence of masses and coupling constants}",
    eprint = "0709.3825",
    archivePrefix = "arXiv",
    primaryClass = "hep-ph",
    reportNumber = "UVIC-TH-07-12, UMN-TH-2618-07, FTPI-MINN-07-28",
    doi = "10.1103/PhysRevD.77.043524",
    journal = "Phys. Rev. D",
    volume = "77",
    pages = "043524",
    year = "2008"
}

@article{Burrage:2017qrf,
    author = "Burrage, Clare and Sakstein, Jeremy",
    title = "{Tests of Chameleon Gravity}",
    eprint = "1709.09071",
    archivePrefix = "arXiv",
    primaryClass = "astro-ph.CO",
    doi = "10.1007/s41114-018-0011-x",
    journal = "Living Rev. Rel.",
    volume = "21",
    number = "1",
    pages = "1",
    year = "2018"
}

@article{Brax:2010gi,
    author = "Brax, Philippe and van de Bruck, Carsten and Davis, Anne-Christine and Shaw, Douglas",
    title = "{The Dilaton and Modified Gravity}",
    eprint = "1005.3735",
    archivePrefix = "arXiv",
    primaryClass = "astro-ph.CO",
    doi = "10.1103/PhysRevD.82.063519",
    journal = "Phys. Rev. D",
    volume = "82",
    pages = "063519",
    year = "2010"
}

@article{Criado:2020jkp,
    author = {Criado, Juan Carlos and Koivunen, Niko and Raidal, Martti and Veerm{\"a}e, Hardi},
    title = "{Dark matter of any spin -- an effective field theory and applications}",
    eprint = "2010.02224",
    archivePrefix = "arXiv",
    primaryClass = "hep-ph",
    doi = "10.1103/PhysRevD.102.125031",
    journal = "Phys. Rev. D",
    volume = "102",
    number = "12",
    pages = "125031",
    year = "2020"
}

@article{Alexander:2020gmv,
    author = "Alexander, Stephon and Jenks, Leah and McDonough, Evan",
    title = "{Higher spin dark matter}",
    eprint = "2010.15125",
    archivePrefix = "arXiv",
    primaryClass = "hep-ph",
    doi = "10.1016/j.physletb.2021.136436",
    journal = "Phys. Lett. B",
    volume = "819",
    pages = "136436",
    year = "2021"
}

@article{Babichev:2016bxi,
    author = {Babichev, Eugeny and Marzola, Luca and Raidal, Martti and Schmidt-May, Angnis and Urban, Federico and Veerm{\"a}e, Hardi and von Strauss, Mikael},
    title = "{Heavy spin-2 Dark Matter}",
    eprint = "1607.03497",
    archivePrefix = "arXiv",
    primaryClass = "hep-th",
    reportNumber = "LPT-ORSAY-16-60",
    doi = "10.1088/1475-7516/2016/09/016",
    journal = "JCAP",
    volume = "09",
    pages = "016",
    year = "2016"
}

@article{Aoki:2019snr,
    author = "Aoki, Katsuki and Mukohyama, Shinji",
    title = "{Ghostfree quadratic curvature theories with massive spin-2 and spin-0 particles}",
    eprint = "1907.09690",
    archivePrefix = "arXiv",
    primaryClass = "hep-th",
    reportNumber = "YITP-19-67, IPMU19-0098",
    doi = "10.1103/PhysRevD.100.064061",
    journal = "Phys. Rev. D",
    volume = "100",
    number = "6",
    pages = "064061",
    year = "2019"
}

@article{Kolb:2023dzp,
    author = "Kolb, Edward W. and Ling, Siyang and Long, Andrew J. and Rosen, Rachel A.",
    title = "{Cosmological gravitational particle production of massive spin-2 particles}",
    eprint = "2302.04390",
    archivePrefix = "arXiv",
    primaryClass = "astro-ph.CO",
    doi = "10.1007/JHEP05(2023)181",
    journal = "JHEP",
    volume = "05",
    pages = "181",
    year = "2023"
}

@article{Babichev:2016hir,
    author = {Babichev, Eugeny and Marzola, Luca and Raidal, Martti and Schmidt-May, Angnis and Urban, Federico and Veerm{\"a}e, Hardi and von Strauss, Mikael},
    title = "{Bigravitational origin of dark matter}",
    eprint = "1604.08564",
    archivePrefix = "arXiv",
    primaryClass = "hep-ph",
    reportNumber = "LPT-ORSAY-16-75",
    doi = "10.1103/PhysRevD.94.084055",
    journal = "Phys. Rev. D",
    volume = "94",
    number = "8",
    pages = "084055",
    year = "2016"
}

@article{Sun:2021yra,
    author = "Sun, Sichun and Yang, Xing-Yu and Zhang, Yun-Long",
    title = "{Pulsar timing residual induced by wideband ultralight dark matter with spin 0,1,2}",
    eprint = "2112.15593",
    archivePrefix = "arXiv",
    primaryClass = "astro-ph.CO",
    doi = "10.1103/PhysRevD.106.066006",
    journal = "Phys. Rev. D",
    volume = "106",
    number = "6",
    pages = "066006",
    year = "2022"
}

@article{Wu:2023dnp,
    author = "Wu, Yu-Mei and Chen, Zu-Cheng and Huang, Qing-Guo",
    title = "{Pulsar timing residual induced by ultralight tensor dark matter}",
    eprint = "2305.08091",
    archivePrefix = "arXiv",
    primaryClass = "hep-ph",
    doi = "10.1088/1475-7516/2023/09/021",
    journal = "JCAP",
    volume = "09",
    pages = "021",
    year = "2023"
}

@article{Xia:2023hov,
    author = "Xia, Zi-Qing and Tang, Tian-Peng and Huang, Xiaoyuan and Yuan, Qiang and Fan, Yi-Zhong",
    title = "{Constraining ultralight dark matter using the Fermi-LAT pulsar timing array}",
    eprint = "2303.17545",
    archivePrefix = "arXiv",
    primaryClass = "astro-ph.HE",
    doi = "10.1103/PhysRevD.107.L121302",
    journal = "Phys. Rev. D",
    volume = "107",
    number = "12",
    pages = "L121302",
    year = "2023"
}

@article{Cai:2024thd,
    author = "Cai, Rong-Gen and Zhang, Jing-Rui and Zhang, Yun-Long",
    title = "{Angular correlation and deformed Hellings-Downs curve from spin-2 ultralight dark matter}",
    eprint = "2402.03984",
    archivePrefix = "arXiv",
    primaryClass = "gr-qc",
    doi = "10.1103/PhysRevD.110.044052",
    journal = "Phys. Rev. D",
    volume = "110",
    number = "4",
    pages = "044052",
    year = "2024"
}

@article{Dias:2023ynv,
    author = "Dias, Oscar J. C. and Lingetti, Giuseppe and Pani, Paolo and Santos, Jorge E.",
    title = "{Black hole superradiant instability for massive spin-2 fields}",
    eprint = "2304.01265",
    archivePrefix = "arXiv",
    primaryClass = "gr-qc",
    doi = "10.1103/PhysRevD.108.L041502",
    journal = "Phys. Rev. D",
    volume = "108",
    number = "4",
    pages = "L041502",
    year = "2023"
}

@article{Blas:2024kps,
    author = "Blas, Diego and Carlton, John and McCabe, Christopher",
    title = "{Massive graviton dark matter searches with long-baseline atom interferometers}",
    eprint = "2412.14282",
    archivePrefix = "arXiv",
    primaryClass = "hep-ph",
    reportNumber = "KCL-PH-TH/2024-69, AION-REPORT/2024-08",
    doi = "10.1103/zxtk-bwnf",
    journal = "Phys. Rev. D",
    volume = "111",
    number = "11",
    pages = "115020",
    year = "2025"
}

@article{Zhang:2025fck,
    author = "Zhang, Jing-Rui and Chen, Ju and Jiao, Heng-Sen and Cai, Rong-Gen and Zhang, Yun-Long",
    title = "{Probing spin-2 ultralight dark matter with space-based gravitational wave detectors in the mHz regime}",
    eprint = "2501.11071",
    archivePrefix = "arXiv",
    primaryClass = "gr-qc",
    doi = "10.1103/cz2w-5cfj",
    journal = "Phys. Rev. D",
    volume = "112",
    number = "6",
    pages = "064030",
    year = "2025"
}

@article{Dubovsky:2004ud,
    author = "Dubovsky, S. L. and Tinyakov, P. G. and Tkachev, I. I.",
    title = "{Massive graviton as a testable cold dark matter candidate}",
    eprint = "hep-th/0411158",
    archivePrefix = "arXiv",
    doi = "10.1103/PhysRevLett.94.181102",
    journal = "Phys. Rev. Lett.",
    volume = "94",
    pages = "181102",
    year = "2005"
}

\iftoggle{endauthorlist}{
  \let\author\myauthor
  \let\affiliation\myaffiliation
  \let\maketitle\mymaketitle
  \title{The LIGO Scientific Collaboration, Virgo Collaboration, and KAGRA Collaboration  }
  \pacs{}
%

\author{A.~G.~Abac\,\orcidlink{0000-0003-4786-2698}}
\affiliation{Max Planck Institute for Gravitational Physics (Albert Einstein Institute), D-14476 Potsdam, Germany}
\author{I.~Abouelfettouh}
\affiliation{LIGO Hanford Observatory, Richland, WA 99352, USA}
\author{F.~Acernese}
\affiliation{Dipartimento di Farmacia, Universit\`a di Salerno, I-84084 Fisciano, Salerno, Italy}
\affiliation{INFN, Sezione di Napoli, I-80126 Napoli, Italy}
\author{K.~Ackley\,\orcidlink{0000-0002-8648-0767}}
\affiliation{University of Warwick, Coventry CV4 7AL, United Kingdom}
\author{C.~Adamcewicz\,\orcidlink{0000-0001-5525-6255}}
\affiliation{OzGrav, School of Physics \& Astronomy, Monash University, Clayton 3800, Victoria, Australia}
\author{S.~Adhicary\,\orcidlink{0009-0004-2101-5428}}
\affiliation{The Pennsylvania State University, University Park, PA 16802, USA}
\author{D.~Adhikari}
\affiliation{Max Planck Institute for Gravitational Physics (Albert Einstein Institute), D-30167 Hannover, Germany}
\affiliation{Leibniz Universit\"{a}t Hannover, D-30167 Hannover, Germany}
\author{N.~Adhikari\,\orcidlink{0000-0002-4559-8427}}
\affiliation{University of Wisconsin-Milwaukee, Milwaukee, WI 53201, USA}
\author{R.~X.~Adhikari\,\orcidlink{0000-0002-5731-5076}}
\affiliation{LIGO Laboratory, California Institute of Technology, Pasadena, CA 91125, USA}
\author{V.~K.~Adkins}
\affiliation{Louisiana State University, Baton Rouge, LA 70803, USA}
% ...
\collaboration{The LIGO Scientific Collaboration, the Virgo Collaboration, and the KAGRA Collaboration}
\author{and \\ J.~Shu\,\orcidlink{0000-0001-6569-403X}}
\affiliation{School of Physics and State Key Laboratory of Nuclear Physics and Technology, Peking University, Beijing 100871, China}
\affiliation{Center for High Energy Physics, Peking University, Yiheyuan Road 5, Haidian District, Beijing 100871, China}
\affiliation{Beijing Laser Acceleration Innovation Center, Huairou, Beiing,101400, China}


  \newpage
  \maketitle
}



\end{document}